\documentclass[reprint, aps, superscriptaddress, pra]{revtex4-2}
\usepackage{amsmath}
\usepackage{amssymb}
\usepackage{amsfonts}
\usepackage{comment}
\usepackage{dsfont}
\usepackage{graphicx}
\usepackage[utf8]{inputenc}
\usepackage[T1]{fontenc}
\usepackage[english]{babel}
\usepackage[usenames,dvipsnames]{xcolor}
\usepackage{mathtools}
\usepackage[caption=false]{subfig}
\usepackage{physics}
\usepackage{graphicx}
\usepackage{dcolumn}
\usepackage{bm}
\usepackage{color}   
\usepackage{hyperref}
\usepackage{ulem}
\hypersetup{
	breaklinks=true,   
    colorlinks=true, 
    citecolor =blue,
    linktoc=all,     
    linkcolor=red,
}

\DeclareUnicodeCharacter{0301}{\'{e}}
\graphicspath{{Figures/}}

\makeatletter
\newcommand*{\rom}[1]{\expandafter\@slowromancap\romannumeral #1@}
\makeatother

\begin{document}
\title{The symmetry in the model of two coupled Kerr oscillators leads to simultaneous multi-photon transitions}
\author{Bogdan Y. Nikitchuk}
\affiliation{Skolkovo Institute of Science and Technology, 121205 Moscow, Russia}
\affiliation{Moscow Institute of Physics and Technology, Dolgoprudny, 141701 Moscow Region, Russia}
\author{Evgeny V. Anikin}
\affiliation{Skolkovo Institute of Science and Technology, 121205 Moscow, Russia}
\author{Natalya S. Maslova} 
\affiliation{
Quantum Technology Centrum, Department of Physics, 
Lomonosov Moscow State University, 119991 Moscow, Russia}
\author{Nikolay A. Gippius}
\affiliation{Skolkovo Institute of Science and Technology, 121205 Moscow, Russia}

\begin{abstract}
We consider the model of two coupled oscillators with Kerr nonlinearities in the rotating-wave approximation. We demonstrate that for a certain set of parameters of the model, the multi-photon transitions occur between many pairs of the oscillator states simultaneously. Also, the position of the multi-photon resonances does not depend on the coupling strength between two oscillators. We prove rigorously that this is a consequence of a certain symmetry of the perturbation theory series for the model. In addition, we analyse the model in the quasi-classical limit by considering the dynamics of the pseudo-angular momentum. We identify the multi-photon transitions with the tunnelling transitions between the degenerate classical trajectories on the Bloch sphere.
\end{abstract}
\maketitle

\section*{Introduction}
For decades, the models consisting of interacting nonlinear oscillator modes attract much attention because of their importance for various 
fundamental concepts of quantum physics, quantum information and nonlinear dynamics. The phenomena present in such models include 
quantum chaos \cite{Adamyan2001}, 
multi-stability \cite{Tadokoro2018}, 
dissipative phase transitions \cite{Zhang2021}, and dynamical tunnelling \cite{Serban2007}. Nonlinear oscillator networks were suggested as a framework for universal quantum computation \cite{Goto2016}, and also quantum nonlinear oscillators were suggested as a tool for non-classical states creation such as squeezed states \cite{PhysRevB.100.035307}, entangled states \cite{Entangled, Entangled3} and cat states \cite{DODONOV1974597}. Also, such models are a fundamental tool to study quantum-classical correspondence \cite{Andersen2020}.

The models of interacting nonlinear oscillators can be realised in various experimental setups including optomechanical systems \cite{Pistolesi2018}, 
trapped ions \cite{Ding2017} and superconducting circuits \cite{Muppalla}. 
The presence of a dynamical bifurcation point was experimentally demonstrated in RF-driven Josephson junctions \cite{Siddiqi2005} and 
transitions between two basins of attraction were observed in a nano-mechanical resonator. 
Also, non-degenerate parametric amplifiers were created based on the Josephson junction arrays \cite{Muppalla}.

Lots of efforts are devoted to studying the nonlinear oscillator systems in the mesoscopic regime \cite{Shirai2018, Andersen2020, Zhang2021}. 
In this regime, it is possible to use the quasi-classical approximation, but the quantum effects are important as well. 
The interplay of complex classical dynamics and quantum effects such as quantum tunnelling leads to new interesting phenomena.
For example, it was shown that in the model of a single oscillator mode with Kerr nonlinearity, 
tunnelling affects the switching rate between the stable states \cite{11}.

The specifics of tunnelling in such systems are determined by the complex structure of the phase space. For the models exhibiting bi- or multi-stability, there exist several different classical trajectories 
with the same energies. This opens way for tunnelling transitions between the corresponding quantum states. However, as the tunnelling transition amplitudes 
are exponentially small, the certain resonant condition should be usually satisfied in order to achieve a non--vanishing tunnelling probability \cite{Serban2007}.
Also, the interconnection between tunnelling and multi-photon transitions was established in many systems \cite{tunMP1,Dykman2005,Anikin2018,tun2MP}. In 
particular, for a single nonlinear oscillator with Kerr nonlinearity, tunnelling between different regions of the phase space is 
equivalent to simultaneous absorption or emission of many oscillator quanta in the quasi-classical limit.

In this work, we consider the model of two coupled nonlinear oscillators in the rotating-wave approximation (RWA) \cite{carm}. The classical limit of this model can be described as the dynamics of the pseudo-angular momentum on a two-dimensional sphere. Among the classical trajectories, there exist ones with equal energies, which makes tunnelling transitions between them possible. We show that tunnelling between such trajectories can be described with 
the perturbation theory in the coupling constant, and tunnelling transition is equivalent to the exchange by many quanta between the oscillators. 
We use high-order perturbation theory to study these processes and account for both resonant and non-resonant contributions. We identify the resonant condition
for tunnelling between different classical trajectories and we prove rigorously that this resonant condition is satisfied simultaneously for many 
pairs of the oscillator states. Also, it is independent of the value of the coupling constant between oscillators, in contrast with many systems demonstrating similar behaviour (such as \cite{OWERRE20151}). This fact is a consequence of an internal symmetry of the system Hamiltonian 
which we establish in all orders of the perturbation theory series. Also, we discuss how the presence of such symmetry modifies 
the energy spectrum and the structure of the eigenstates.

These results could have applications for quantum information processing and quantum state manipulation, in particular, the 
generation of the entangled states of two modes. 
We believe that the predicted features can be observed in systems with high–quality oscillator modes with a pronounced Kerr nonlinearity, such as the plasmon modes of Josephson junction arrays or the phonon modes of trapped ion ensembles. Both mentioned systems represent a natural realization of a system of coupled quantum nonlinear oscillators well isolated from the environment
Also, the considered model is closely related to some models \cite{DTC1, DTC2} of dissipative time crystals (DTC) \cite{DTCrev1, DTCrev2}. 
The results of this work could be useful to study the quantum effects in the DTC. In addition, the discovered symmetry extends the range of known symmetries in quantum--optical systems \cite{sym11101310, sym2, sym3, sym4}. 

\onecolumngrid

\section{Multi-photon transitions between two coupled nonlinear oscillators}
\label{sec:the_model}
We consider the model of two oscillators with linear coupling and
Kerr nonlinearities. Let the frequencies of the oscillators be $\omega_{1,2}$,
and the Kerr shifts per oscillators 
quanta be $\alpha_{1,2}$. When the detuning between the oscillators
$\Delta = \omega_2 - \omega_1$ is much smaller than the oscillators
frequencies, one can neglect the counter-rotating terms in coupling (RWA) \cite{carm}, and 
the model Hamiltonian reads
\begin{equation}
\hat{H}=\omega_1 \hat{a}^{\dagger} \hat{a}+\omega_2\hat{b}^{\dagger} \hat{b} +
\frac{\alpha_{1}}{2}\left(\hat{a}^{\dagger} \hat{a} \right)^{2}+\frac{\alpha_{2}}{2} \left(\hat{b}^{\dagger} \hat{b} \right)^{2}+ 
g\left(\hat{a}^{\dagger} \hat{b}+\hat{b}^{\dagger} \hat{a}\right), \label{ham}
\end{equation} 
where $g$ is the coupling constant between the oscillators.
As the counter-rotating terms are not present in \eqref{ham},
this Hamiltonian commutes with the total number of quanta operator
$\hat{N} = \hat{a}^{\dagger} \hat{a} + \hat{b}^{\dagger} \hat{b}$ (we will denote its eigenvalues as $N$).
Therefore, the total Hilbert space $\mathcal{H}$ of the considered model \eqref{ham}
splits into the direct sum of invariant Hilbert subspaces each corresponding 
to $N$ quanta in the oscillators: $\mathcal{H} =  \bigoplus\limits_{N=0}^{\infty} \mathcal{H}_N$, and 
also $\hat{H} = \sum_N \hat{H}_N$. 
The Hamiltonians $\hat{H}_N$ which act on the subspace $\mathcal{H}_N$ can be written with help of bra-ket notation as
\begin{multline}
    \label{eq:ham_N}
      \hat{H}_{N}=\frac{1}{2}\left(\alpha_{1}+\alpha_{2}\right) 
      \sum_{n=0}^{N}  n(n - \mu_N)|n, N-n\rangle\langle n, N-n|+ \\
      g \sum_{n=0}^{N}  \sqrt{n(N-n+1)}\Big(|n-1, N-n+1\rangle\langle n, N-n|+
|n,N-n\rangle\langle n-1, N-n+1|\Big)+ \omega_2 N + \frac{\alpha_2 N^2}{2},
\end{multline}
where $|n_a, n_b\rangle = \ket{n_a} \otimes \ket{n_b}$ are the oscillators Fock states ($n_a$ and $n_b$ are the numbers of quanta in modes $a$ and $b$ respectively), and
\begin{equation} 
\label{eq:mu} 
\mu_N = 2 \left( \Delta + \alpha_2 N \right) / \left( \alpha_1 + \alpha_2 \right), 
\end{equation}

The model \eqref{ham} is closely related to the model of a single Kerr oscillator \cite{Drummond} 
in the classical external field. Namely, in the limit $N\to\infty$, 
$g\sqrt{N} = \mathrm{const}$, the Hamiltonian $\hat{H}_N$ approaches the Hamiltonian of a single 
Kerr oscillator in the classical external field 
with nonlinearity $\alpha_1~+~\alpha_2$, detuning $\alpha\mu_N/2$, and the external
field amplitude $~g\sqrt{N}$. This can be seen from Eq.~\eqref{eq:ham_N}.

\begin{figure}[h!]
\center{\includegraphics[width = 17 cm]{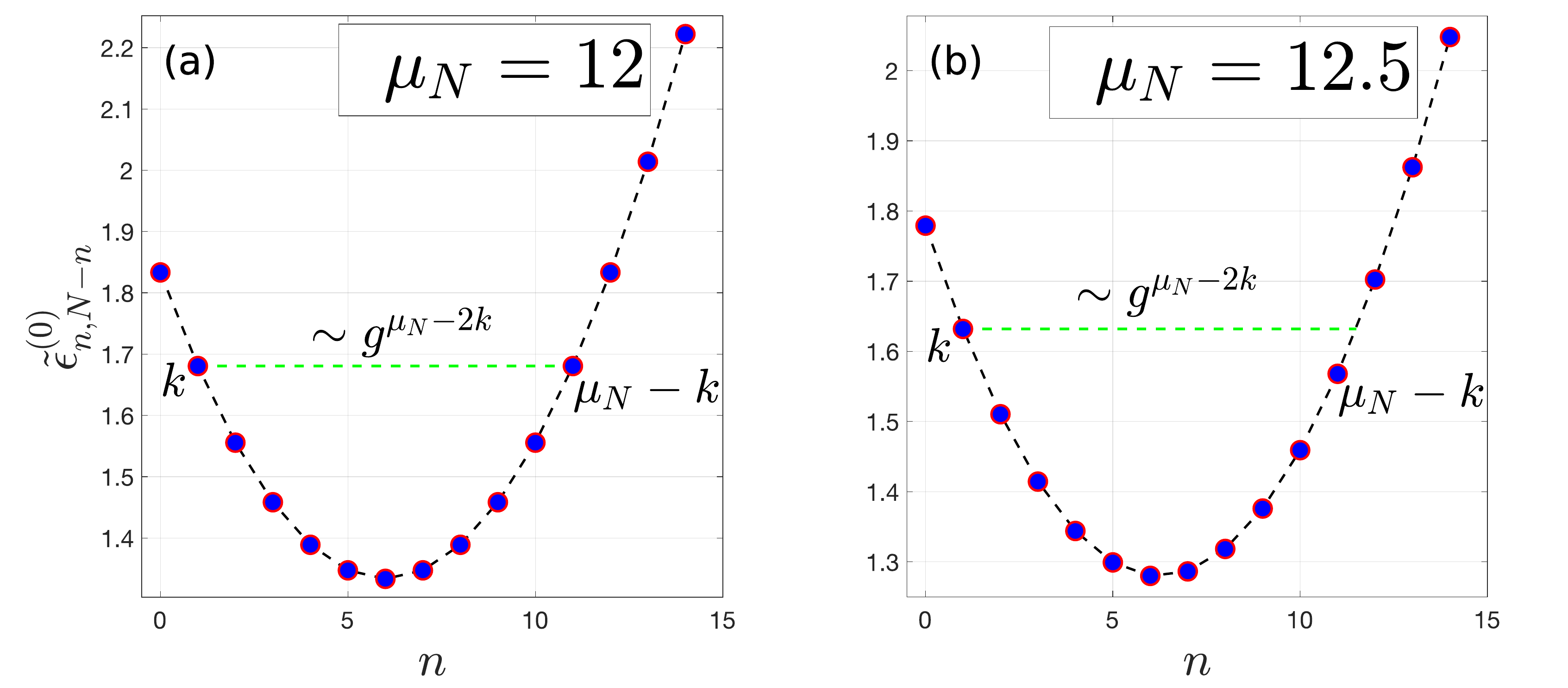}} 
\caption{The dimensionless eigenenergies $\tilde{\epsilon}^{(0)}_{n, N-n}=(\alpha_1 + \alpha_2) \epsilon^{(0)}_{n,N-n} / (\Delta + \alpha_2 N)^2$ 
  (with $\epsilon$ denoting the unnormalized eigenenergies)
of the Hamiltonian \eqref{ham} at $g=0$, $N=14$, 
$\alpha_2 / \alpha_1 = 1.1$, and (a) $\mu_N = 12$ and (b) $\mu_N = 12.5$
as functions of $n$ for fixed $N$.}
\label{fig:parabola}
\end{figure}

In the following, we will focus on the case of relatively weak coupling.
Moreover, in Sections \ref{sec:symmetry_proof} and \ref{sec:high_ord}, 
we will use the perturbation theory in $g$ as a theoretical tool. 
Because of that, let us first consider the model without perturbation. 
In this case, the Hamiltonian \eqref{ham} commutes both with $a^\dagger a$ and $b^\dagger b$, and its eigenstates are the Fock states $|n_a, n_b\rangle$ of the oscillators. The energy of each state $|n_a, n_b\rangle$ can be found easily from Eq.~\eqref{ham}

\begin{equation} 
\label{spectrun_no_coup} 
\epsilon_{n_a,n_b}^{(0)} = 
\frac{\alpha_1}{2}n_a^2 + \frac{\alpha_2}{2}n_b^2 + \omega_1 n_a + \omega_2 n_b.
\end{equation}

The perturbation operator allows the oscillator to exchange quanta
between each other. Namely, the perturbation directly couples $|n_a, n_b\rangle$
and $|n_a\pm 1, n_b\mp 1\rangle$. 
Because of that, the states $|n_a,n_b\rangle$ and $|n_a + k,n_a - k\rangle$
also become coupled in the order $k$ by perturbation, 
and the transition amplitude between them is proportional to $g^{k}$. It turns out
that at certain values of the detuning between the oscillators, such
transitions (we will call them multi-photon) become resonant, and
even at small couplings, the oscillators can exchange by $k$ quanta
simultaneously between each other.

This occurs because the unperturbed energies of the oscillators 
Fock states $|n,N-n\rangle$ in each subspace corresponding to $N$ quanta have a parabolic dependence on $n$. The way how a sequence of the values $\epsilon_{n,N}^{(0)}$ 
are arranged on a parabola depends on the value of 
the parameter $\mu_N$. 

One can easily see that at integer $\mu_N=m\in\mathbb{Z}$, many 
energies split into pairs with equal
values: $\epsilon^{(0)}_{n,N-n} = \epsilon^{(0)}_{m-n,N-m+n}$ for $n = 0, \dots m$ (see Fig.~\ref{fig:parabola}(a)). In contrast, for non--integer 
$\mu_N$ the unperturbed energies are non-degenerate (see Fig.~\ref{fig:parabola}(b)).
Also, for the case $\alpha_1 = \alpha_2$, the resonance condition is simultaneously satisfied for all $N$.

When $\mu_N$ is close to an integer, the degeneracy makes multi-photon transitions possible between the states $|n,N-n\rangle$ 
and $|m-n,~N-m+n\rangle$ when the perturbation is turned on. In other
words, the integer values of $\mu_N$ at small $g$ correspond to multi-photon resonances between  $|n,N-n\rangle$ and $|m-n, N-m+n\rangle$.

\begin{figure}[h!]
\center{\includegraphics[width=12.75 cm]{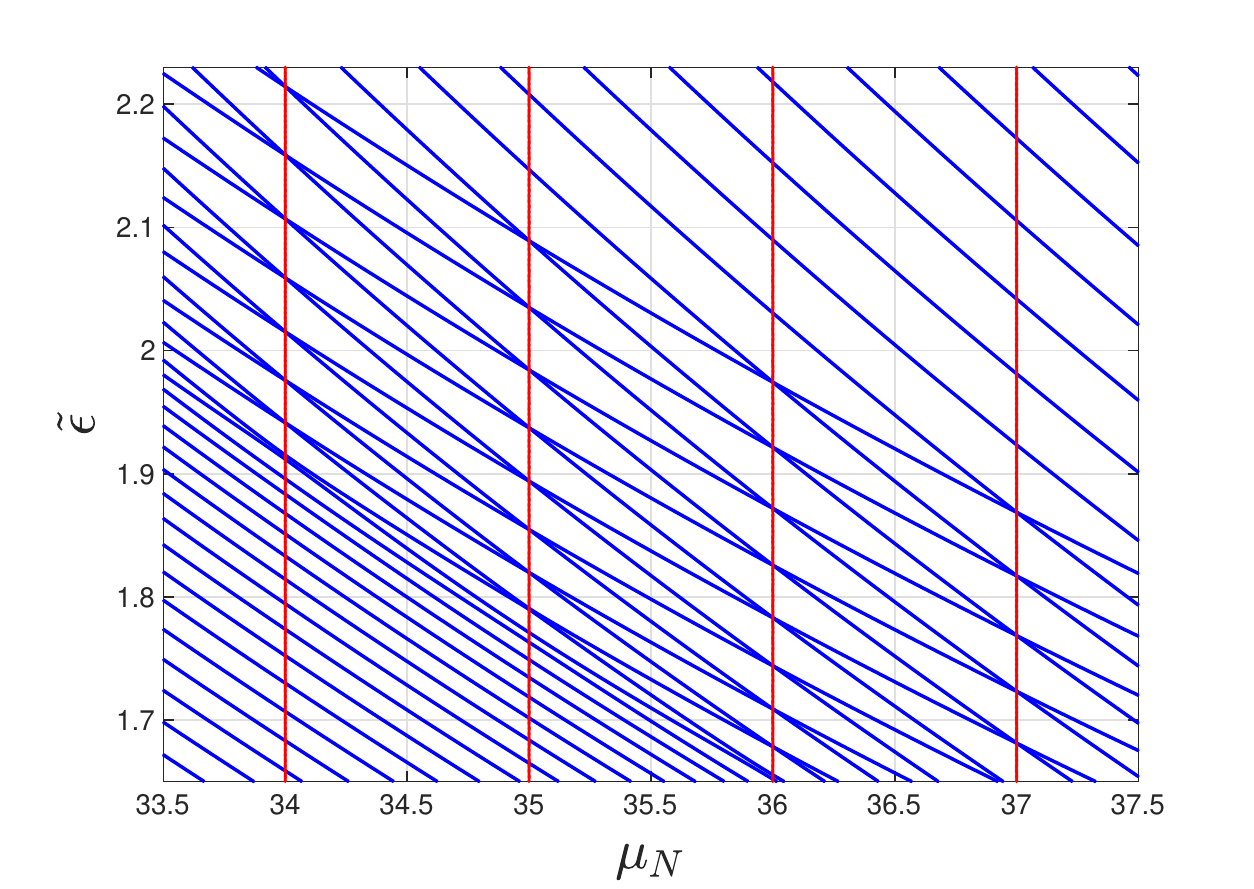}}
\caption{
The dimensionless eigenenergies $\tilde{\epsilon} = (\alpha_1 + \alpha_2) \epsilon/ (\Delta + \alpha_2 N)^2$ 
  (with $\epsilon$ denoting the unnormalized eigenenergies)
of the Hamiltonian \eqref{ham} corresponding to the subspace with $N = 46$ quanta at $\alpha_2 / \alpha_1 = 1.5$, 
$\Delta = \mu_N (\alpha_1 + \alpha_2)/2 - \alpha_2 N$, and $g = 0.32g_\mathrm{crit}$ (see Section~\ref{sec:classical_limit} for the definition of $g_\mathrm{crit}$) 
are shown as functions of $\mu_N$. 
All the anti-crossing points are located on the vertical lines corresponding to integer values of $\mu_N$.}
\label{fig:spectra}
\end{figure}

However, the perturbation not only leads to multi-photon transitions between 
$|n, N - n\rangle$ and $|m - n, N - m + n\rangle$, but also induces 
the energy shifts of these levels due to the 
non-resonant coupling with other levels. 
In general, these shifts could lift the position of 
the multi-photon resonances from 
exactly $\mu_N \in \mathbb{Z}$, and to make them dependent on $g$. The surprising feature of the 
considered model is that these energy shifts are equal for every 
degenerate pair $|n, N - n\rangle$ and $|m - n, N - m + n\rangle$  (see Fig.~\ref{fig:spectra}). 
Because of that, simultaneous multi-photon resonances 
between many level pairs occur exactly at integer values of $\mu_N$ 
even at relatively large values of  $g$. 

In the absence of degeneracy in the unperturbed model (for the case of non-integer $\mu_N$), 
weak coupling between the oscillators leads only to a small $O(g)$ perturbation of the Fock states of
the system: the eigenstates remain close to Fock states. This is not the case 
when the multi-photon resonance condition is satisfied. Due to the reasons indicated above, even
at small couplings the structure of the Hamiltonian eigenstates and the temporal dynamics of the system wave function for integer $\mu_N$ qualitatively differ from the case of non-interacting
oscillators. 

\begin{figure}[h!]
\center{\includegraphics[width=17cm]{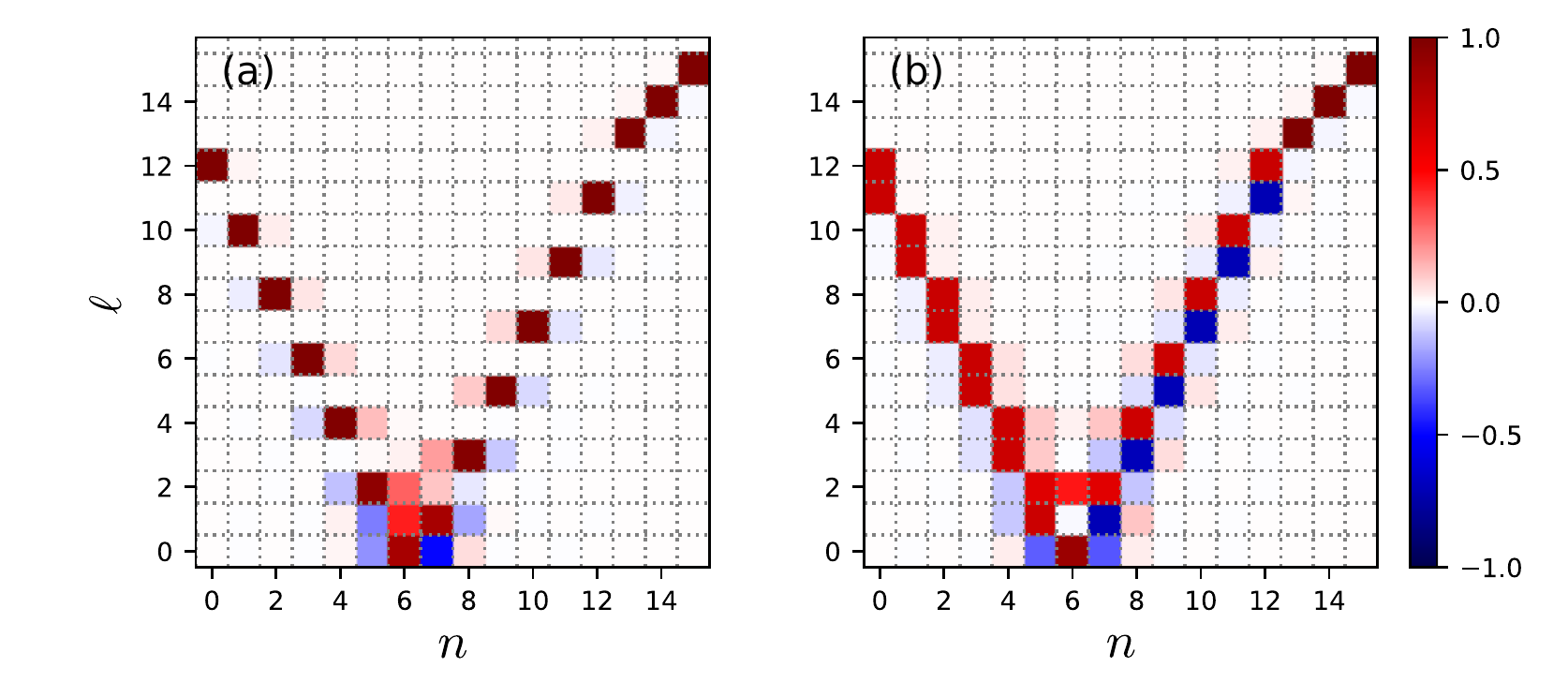}}
\caption{The expansion coefficients $c_{\ell, n}$ of the eigenstates of the Hamiltonian \eqref{ham} 
at $N = 12$, $\alpha_2/\alpha_1 = 1$,  $g/\alpha_1 \approx 0.03$, (a) $\Delta/ \alpha_1 = 6.25$ and (b) $\Delta/ \alpha_1 = 6.5$
are shown.
For each pair $n, \ell$, a unit square centered at the corresponding point of the figure is drawn. The color of the square indicates
the magnitude of $c_{n,\ell}$ according to the color bar.}
\label{fig:bars}
\end{figure}

In this case, the leading-order contributions to the wave functions 
are symmetric and antisymmetric superpositions of the Fock 
states $|n,N-n\rangle$ and $|m-n, N-m+n\rangle$. For every $n = 0,\dots, m$,
\begin{equation}
\label{eigenstates}
|\psi_{n,m-n}^\pm\rangle \sim \frac{1}{\sqrt{2}} \big(|n,N-n\rangle \pm |m-n, N-m+n\rangle \big) + O(g).
\end{equation}
This can be seen from the results of 
the numerical diagonalization of the Hamiltonian \eqref{ham}.
We find the eigenstates of the Hamiltonian \eqref{ham} 
in the Fock basis in the form
$\ket{\psi_{\ell}}~=~\sum_n c_{\ell, n}~\ket{n, N-n}$.
The eigenstates are sorted in the 
ascending order according to their eigenenergies. 
Then, we plot the matrices of the expansion coefficients $c_{\ell, n}$ 
in Fig.~\ref{fig:bars} for the cases of 
integer (Fig.~\ref{fig:bars}a) and non--integer $\mu_N$ 
(Fig.~\ref{fig:bars}b). For each eigenstates shown in Fig.~\ref{fig:bars}a,
there is a single Fock state which has dominant contribution, 
whereas the eigenstates
in Fig.~\ref{fig:bars}b contain equal contributions from two Fock states 
as in Eq.~\eqref{eigenstates}.

\begin{figure}[h!]
  \center{\includegraphics[width = 12.75 cm]{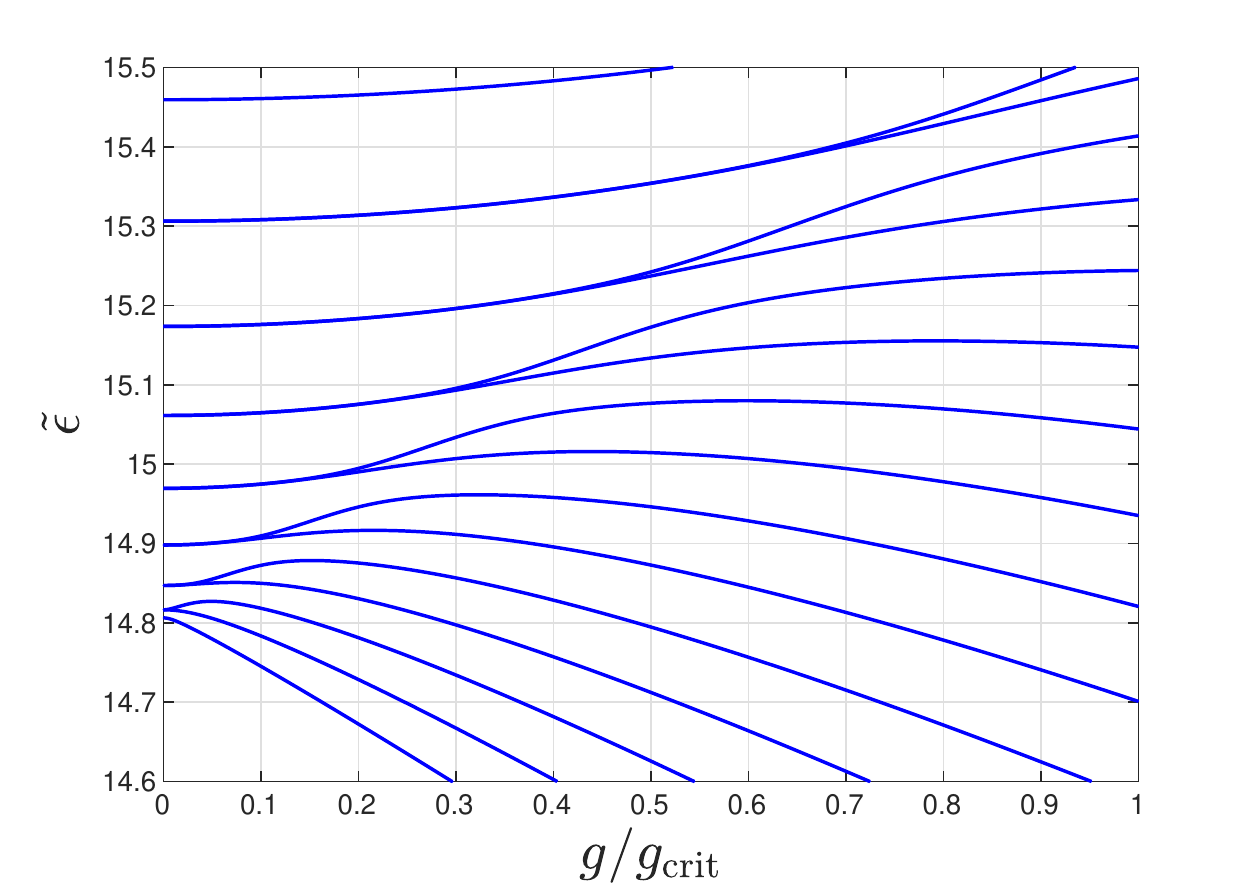}}
\caption{The dimensionless eigenenergies $\tilde{\epsilon} = (\alpha_1 + \alpha_2) \epsilon / (\Delta + \alpha_2 N)^2$ 
  of the Hamiltonian \eqref{ham} (with $\epsilon$ denoting the eigenenergies)
as functions of the coupling $g$ for $\mu_N = 14$, 
$\alpha_2 / \alpha_1 = 1.5$, $N = 50$ and $\Delta = \mu_N (\alpha_1 + \alpha_2)/2 - \alpha_2 N$. 
For the definition of $g_\text{crit}$, see Section~\ref{sec:classical_limit}.
}
\label{fig:splitting}
\end{figure}

The energy difference between the states $|\psi_{n,m-n}^\pm\rangle$ 
exhibiting the anti-crossing is determined by the multi-photon transition
amplitude $\omega^R_{n,m-n} \propto g^{m-2n}$ (see Fig.~\ref{fig:splitting})
\begin{equation} 
  \label{eq_split}
  \epsilon_{n,m-n}^+ - \epsilon_{n,m-n}^- = 2\omega^R_{n,m-n} + O\left(g^{m-2n+1}\right).
\end{equation}

In addition, from Eq.~\eqref{eigenstates} and \eqref{eq_split}, for the initial state: $\ket{\psi(0)}~=~\sum_{n} c_n \ket{n, N-n}$, one can find (at integer $\mu_N = m$) the following approximate solutions of the non-stationary Schrodinger equation
\begin{equation} \label{eq_evol}
|\psi  (t)\rangle  =  \sum\limits_{n}c_n e^{-\frac{i}{2}\left(\epsilon_{n,m-n}^+ + \epsilon_{n,m-n}^-\right)t}
 \Big[\cos{(\omega^R_{n,m-n} t)} |n,N-n\rangle - 
   i \sin{(\omega^R_{n,m-n} t)} |m-n,N-m+n\rangle \Big]. 
\end{equation}
As can be seen, the system exhibits multi-photon Rabi oscillations between 
many pairs of the Fock states simultaneously.

\begin{figure}[h!]
\center{\includegraphics[width=\textwidth]{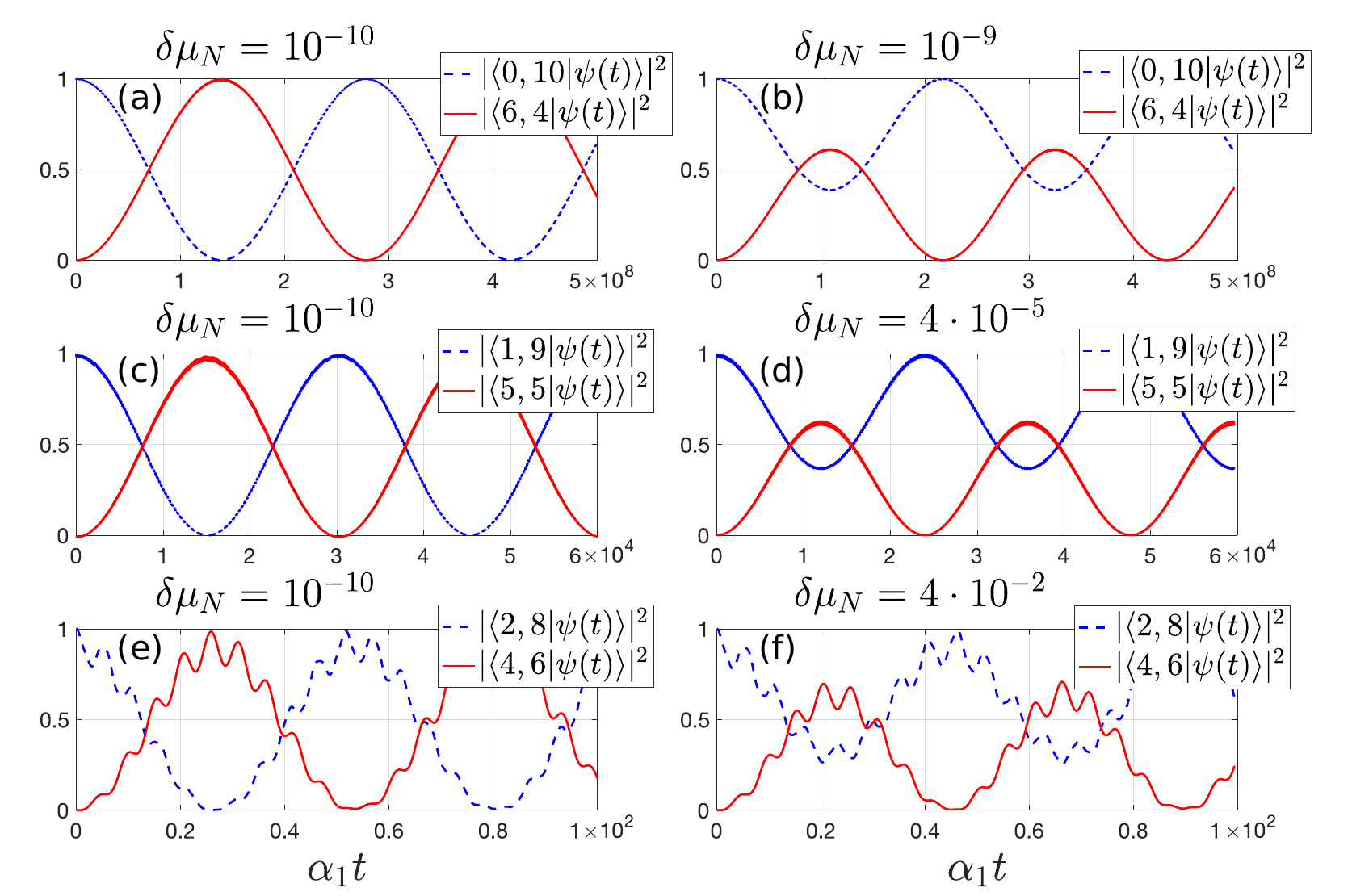}}
\caption{For the wave function $\ket{\psi(t)}$ solving the Schrodinger equation with the initial condition $\ket{\psi(0)} = \ket{n,N-n}$, 
  the projections on $\ket{n, N-n }$ and $\ket{m-n, N-m+n}$ are shown for (a, b) $n = 0$, (c, d) $n = 1$, (e, f) $n = 2$, $N = 10$,
  $\alpha_2 / \alpha_1 = 1$, $\Delta = \mu_N (\alpha_1 + \alpha_2)/2 - \alpha_2 N \approx -4$, 
  $g / \alpha_1 = 0.05$, $\mu_N = m + \delta \mu_N$, $m = 6$, and different $\delta \mu_N$.
  }
\label{fig:compp}
\end{figure}

In Fig.~\ref{fig:compp}, we show the numerical 
solution of the Schrodinger equation for different Fock states taken 
as initial conditions. We plotted the squares of modulus of the overlappings between the wave function $\ket{\psi(t)}$ and the bra-states $\bra{n,N-n}$, $\bra{m-n,N-m+n}$ for different $n$. This result is in agreement with Eq.~$\eqref{eq_evol}$ for integer $\mu_N$. 
Also, for the case of non-integer 
$\mu_N = m + \delta \mu_N$ with sufficiently small $\delta\mu_N$, the amplitude of multi-photon Rabi oscillations 
between $|n, N-n\rangle$ and $|m-n, N-m+n\rangle$ decreases with increasing $\delta\mu_N$. 
They completely vanish when $(\alpha_1+\alpha_2)\delta\mu_N \gg \omega_{n,m-n}^R$ (see left and right panels of Fig.~\ref{fig:compp}).
In addition, there are corrections to Eq.~\eqref{eq_evol} 
which come from the non-resonant contributions of the adjacent Fock states. 
They lead to the additional modulations of the Rabi oscillations 
with the amplitude $\sim g/\Delta$ and become more pronounced with 
increasing $n$ due to the dependence of the transition matrix elements and the energy differences on $n$. They can be seen on the lower panels of Fig.~\ref{fig:compp}. For other panels, they are also present but not resolved on the plots.

Also, let us briefly discuss the effect of weak dissipation on the considered multi-photon Rabi oscillations. In the presence of dissipation, each eigenstate of the system of coupled oscillators obtains a finite lifetime having the same order of magnitude as the decay rates $\gamma_a$, $\gamma_b$ of the individual bosonic modes \cite{Breuer2007}. The condition necessary to observe the Rabi oscillations between the states 
$|n,N-n\rangle$ and $|m-n,N-m+n\rangle$ is that the lifetime of 
the eigenstates  $|\psi_{n,m-n}^\pm\rangle$ is much longer than the period of Rabi oscillations $\omega_{n,m-n}^R$, which leads to conditions $\gamma_{a,b} \ll \omega_{n,m-n}^R$. The requirements to observe the multi-photon transitions between two oscillator states become increasingly strict with the increasing order of the multi-photon transition. Thus, in the presence of dissipation, it is realistic to observe several transitions between the states $|n,N-n\rangle$ and $|m-n,N-m+n\rangle$ with the lowest values of $|m-2n|$.

The independence of the anti-crossings position on the value of $g$ can have practical applications for the measurements of the Kerr coefficients 
of the oscillators.
More precisely, assume the detuning~$\Delta$ between the oscillators is scanned to measure some observable. Often, the multi-photon resonances 
lead to peaks/dips in the corresponding dependencies on $\Delta$. According to the above considerations, the positions of these peaks correspond to 
the integer multiples of the total nonlinearity independently of the coupling constant. This could be used for the measurement of $\alpha_1 + \alpha_2$.

In addition, the multi-photon transitions provide a way to prepare the 
entangled states of two oscillators. For example, according to Eq.~\eqref{eq_evol}, the initial Fock state $\ket{n, N-n}$ 
evolves into an entangled state at the proper choice of the interaction time (when $\omega^R_{n,m-n} t \sim \pi / 4$).

\section{The proof of the symmetry properties}
\label{sec:symmetry_proof}
In this section, we provide a proof (on the physical level of rigour) of the presence of many simultaneous anti-crossings at $\mu_N \in \mathds{Z}$ for the model \eqref{ham}.

Let us consider the coupling term between the oscillators as a perturbation. As commented above, at integer $\mu_N = m$, the unperturbed energies split 
into $\lfloor m/2 \rfloor$ pairs of degenerate levels. The perturbation leads to two effects: the shifts of the energy levels due to the non-resonant couplings, and the multi-photon transitions between the degenerate levels. The consistent analysis of both of the effects requires the usage of 
high-order perturbation theory.

Because of the degeneracies, one can not use directly the 
non-degenerate perturbation theory series in $g$ for the 
system energies. However, at non-integer $\mu_N$ one can define the 
energies $\epsilon_n(g)$ which correspond to the eigenstates evolving 
from the $n$-th Fock state $\ket{n, N-n}$ after the adiabatic 
switching of the perturbation. These energies (at non-integer $\mu_N$) can
be found from non-degenerate perturbation theory as power series 
\begin{equation}
  \label{eq:nondegen_pert_exp}
  \epsilon_n(g) = \epsilon_n^{(0)} + \sum\limits_{k = 1}^{\infty} \epsilon_n^{(k)} g^k.
\end{equation}
(The expressions for the second and fourth order corrections could be found in~Appendix~\ref{Appendix_corrections}).

To some extent, this expansion is valid for $\mu_N \in \mathds{Z}$ as
well. Although at $\mu_N\in\mathds{Z}$ the states $|n, N-n\rangle$ and 
$|m - n, N-m+n\rangle$ are degenerate, the perturbation couples them only
in the order $|m-2n|$. 
Therefore, up to this order of the perturbation theory, 
they can be treated as non-degenerate ones. 
Also, the perturbation theory coefficients $\epsilon_n^{(k)}$ are rational functions of $n$ and $\mu_N$, therefore, they are well--behaved at 
$\mu_N\in \mathds{Z}$. So, one can use the 
expansion \eqref{eq:nondegen_pert_exp} up to the order $|m-2n|$ in the 
degenerate case.

Surprisingly, the following identity is valid for the perturbation 
theory corrections at $\mu_N = m \in \mathds{Z}$:
\begin{equation}
  \label{eq:corrections_equality}
  \epsilon_n^{(k)} = \epsilon_{\mu_N-n}^{(k)}.
\end{equation}
Therefore, the degeneracy between the levels $|n,N-n\rangle$ and
$|m-n,N-n+m\rangle$ is not lifted in several lowest orders of the
perturbation theory until the order $|m-2n|$. This explains the 
Eq.~\eqref{eq_split} and the results of the numerical diagonalization. 

Strictly speaking, the identity \eqref{eq:corrections_equality} makes sense only 
for the case of the integer $\mu_N = m$. However, as the corrections 
$\epsilon_{n}^{(k)}$ are rational functions of $n$ and $\mu_N$, they 
can be analytically continued to the case of arbitrary real $n$ and
$\mu_N$. We will prove \eqref{eq:corrections_equality} 
for these analytical continuations.

The analytical continuation of 
$\epsilon_{n}^{(k)}$ for the case of non-integer $n$ can 
be made in a natural way for all $k$. For that, the whole energy 
spectrum $\epsilon_{n}(g)$ should be continued to the case of non-integer 
$n$.
To do this, let us assume that the Hamiltonian (\ref{ham}) acts on the space of all possible real <<numbers of quanta>> $\nu$ with formally defined matrix elements as
\begin{equation} \label{op}
\langle\nu, N - \nu-1| \hat{a} | \nu+1, N - \nu - 1\rangle=  
\left\langle\nu+1, N - \nu \left|\hat{a}^{\dagger}\right| \nu, N - \nu\right\rangle=\sqrt{\nu+1},
\end{equation}
and analogously for $\hat{b}$ and $\hat{b}^\dag$.
As one can see, in the case of an integer $\nu$, 
the definition of (\ref{op}) reduces to the usual action
of the bosonic creation and annihilation operators in the Fock space. 
We shall note that $\hat{a}$~and~$\hat{a}^\dag$
are no longer Hermitian conjugate with each other 
for the case of non-integer $\nu$, 
so the Hamiltonian becomes non-Hermitian. 
However, one can check that the wave functions remain normalizable
even for the Hamiltonian at the non-integer $\nu$.

The extended Hamiltonian acts invariantly on each of the subspaces 
$\mathcal{V}_\nu =\{|\sigma, N - \sigma \rangle: \sigma - \nu \in \mathds{Z}\}$. 
(The subspace $\mathcal{V}_\nu$ consists of the ket vectors $|\sigma\rangle$ such as $\sigma - \nu \in \mathds{Z}$.)
Also, the subspaces of states $|\nu, N - \nu \rangle$ with negative 
integer $\nu$, 
$\nu \in [0, N]$ 
and integer $\nu > N$ are 
decoupled from each other.
So, the extended Hamiltonian for all real $\nu$ acts exactly as the original Hamiltonian \eqref{ham} on the subspace of $|n, N - n\rangle$, $n\in [0, N]$.
We emphasize that the extension to the non--integer numbers of quanta should be taken as an auxiliary tool for the proof. Because, as we mentioned above, we are interested in the case of an integer $\nu$ which corresponds to the initial Hermitian Hamiltonian. 

Until the end of this section, we will work in 
the subspace of a fixed number of quanta $N$. We will 
denote $\ket{\nu, N - \nu}$ as $\ket{\nu}$ for brevity.

Let us define the action of the extended Hamiltonian on 
$\mathcal{V}_\nu$ as $\mathcal{H}_{\nu, N}$. The operator 
$\mathcal{H}_{\nu, N}$ can be written in terms of the states $|\nu\rangle$ as
\begin{equation}
  \label{eq:extended_ham}
\hat{\mathcal{H}}_{\nu, N}=\frac{1}{2}\left(\alpha_{1}+\alpha_{2}\right) \sum\limits_{\sigma-\nu \in \mathbb{Z}} \sigma(\sigma-\mu_N)|\sigma\rangle\langle\sigma|+ 
g \sum\limits_{\sigma-\nu \in \mathbb{Z}} \sqrt{\sigma(N-\sigma+1)}(|\sigma-1\rangle\langle\sigma|+| \sigma\rangle\langle\sigma-1|). 
\end{equation}
To prove the symmetry of the perturbation theory corrections to the energy, 
we will show the exact symmetry of the eigenstates of the extended 
Hamiltonian with respect to the replacement $\nu \to \mu_N - \nu$. Namely, 
we will prove that $\mathcal{H}_{\nu, N}$ and 
$\mathcal{H}_{\mu_N-\nu, N}$ have the same eigenenergies $\epsilon_{\nu}(g)$ and $\epsilon_{\mu_N - \nu}(g)$ respectively as a functions of $g$. 
For that, we will show that there exists a linear operator $\mathcal{T}$ such \\
\begin{equation}
  \label{eq:symmetry}
  \hat{\mathcal{H}}_{\nu, N}=\mathcal{T} I \hat{\mathcal{H}}_{\mu_N-\nu, N} I^{-1} \mathcal{T}^{-1},
\end{equation}
where $I$ is the isomorphism between the vector spaces spanned by vectors $\ket{\nu}$ and $\ket{\mu_N - \nu}$ respectively.

The existence of the operator $\mathcal{T}$ proves that the operators $\mathcal{H}_{\nu, N}$ and $\mathcal{H}_{\mu_N-\nu, N}$ have identical spectra at any $g$, which means that 
\begin{equation}
  \label{eq:full_energies_equality}
  \epsilon_{\nu}(g) = \epsilon_{\mu_N-{\nu}}(g), \qquad 2 \nu-\mu_N \notin \mathbb{Z}.
\end{equation}

In~Appendix~\ref{D}, we explicitly construct the operator 
$\mathcal{T}$ and show that 
it can be expressed in the following form: $\mathcal{T} = UTV^{-1}$, where
\begin{equation} \label{eq:operat}
\begin{gathered}
T = \left( \mathds{1} - \frac{2g}{\left(\alpha_{1}+\alpha_{2}\right) } \sum\limits_{\sigma - \nu \in \mathbb{Z}} \ket{\sigma} \bra{\sigma + 1} \right)^{- (N-\mu_N)}, \\
U=\sum\limits_{\sigma} \sqrt{\frac{\Gamma(\sigma+1)}{\Gamma(N-\sigma+1)}}|\sigma\rangle\langle\sigma|, \quad
V=\sum\limits_{\sigma} \sqrt{\frac{\Gamma\left(\mu_{N}-\sigma+1\right)}{\Gamma\left(N-\mu_{N}+\sigma+1\right)}}|\sigma\rangle\langle\sigma|.
\end{gathered}
\end{equation}

After we proved the equality of the energies 
\eqref{eq:full_energies_equality},
let us turn back to the physically meaningful case of
the integer $\nu$ and $\mu_N = m$. As we mentioned before, the energy $\epsilon_{\nu}(g)$ can be decomposed in perturbation theory series~of~$g$ (compare 
with Eq.~\eqref{eq:nondegen_pert_exp}):
\begin{equation} \label{decomposition}
\epsilon_{\nu}(g)=\frac{1}{2}(\alpha_1 + \alpha_2) \nu(\nu-\mu_N)+\sum_{k=1}^{\infty} \epsilon_{\nu}^{(k)} g^{k}.
\end{equation}
From the equality \eqref{eq:full_energies_equality} valid for all orders in
$g$,
one concludes that this holds for each term of the perturbation series:
\begin{equation} 
 \label{equalities}
 \epsilon_{\nu}^{(k)} = \epsilon_{\mu_N-\nu}^{(k)} \quad \forall k,
 \quad 2 \nu-\mu_N \notin \mathbb{Z}.
\end{equation}
From Eq.~\eqref{equalities}, the 
equality \eqref{eq:corrections_equality} can be derived easily.
The $k$-th order corrections $\epsilon_{\nu}^{(k)} $ 
are rational functions of $\mu_N$ and $\nu$ and have no singularities 
at $2\nu - \mu_N \in \mathds{Z}$ unless $k \geqslant 2|\mu_N-2\nu|$ (we prove it rigorously in the next sections). 
Therefore, the equality (\ref{equalities}) holds even 
for the case of an integer $2\nu-\mu_N$, and also for 
$\nu, \mu_N \in \mathds{Z}$. This concludes the proof of the equality~\eqref{eq:corrections_equality}.

We should note that the analogous properties hold for the model of a single nonlinear mode with Kerr nonlinearity
driven by the classical external field (recovered as a limit of the considered model at $N \to \infty$). For this model, the equality of the perturbation
theory corrections was checked \cite{PhysRevA.38.1349} for several lowest orders, and 
the sketch of the proof was given previously \cite{Anikin2019}.

\section{High-order perturbation theory for the degenerate energy levels}
\label{sec:high_ord}
We have stated that non-degenerate perturbation theory corrections are symmetric with respect to replacement $n \to \mu_N - n$ in all orders. However, additional arguments are needed to relate this 
result to the physical case of $\mu_N \in \mathds{Z}$. On one hand, due to the presence of degeneracy in the energy spectrum, one should use the degenerate perturbation theory. However, on the other hand, the off-diagonal matrix elements in the secular equation occur only in the $|m-2n|$-th order of perturbation theory. To demonstrate that until the $|m-2n|$-th order one can use the non-degenerate perturbation theory and to calculate the multi-photon amplitude via degenerate perturbation theory for higher orders than $|m-2n|$, it is convenient to apply Green's function formalism. 
Namely, we consider the operator of Green's function defined as
\begin{equation} \label{Green_1}
    \hat{G}(\omega) = \left[\omega - \hat{H}\right]^{-1}.
\end{equation}

If the spectrum of the problem $ \sigma(\hat{H}) = \{\epsilon_n, \ket{\psi_n} \}$ is known, Eq.~\eqref{Green_1} can be rewritten as follows
\begin{equation} \label{def_Green}
\hat{G}(\omega)=\sum_{n} \frac{\left|\psi_{n}\right\rangle\left\langle\psi_{n}\right|}{\omega-\epsilon_{n}}.
\end{equation}
Thus, eigenenergies are poles of Green's function and eigenfunctions could be calculated from residues of $\hat{G}$. The matrix element $G_{n,n}$ 
for each Fock state $|n\rangle$
can be calculated from the Dyson equation and reads
\begin{equation}
G_{n, n}(\omega)=\frac{1}{\omega-\epsilon_{n}^{(0)}-\Sigma_{n}(\omega)},
\end{equation}
where $\Sigma_n$ is the self-energy term, defined as the sum of all diagrams which
start and finish at $n$ and do not contain the Green's function $G_{n,n}$. When $\Sigma_n$ is the regular function in the vicinity of $\omega = \epsilon_n^{(0)}$, the position of the pole of $G_{n,n}$ can be found from the equation $G_{n, n}^{-1}(\omega)=0$ as power series in $g$ which coincides with the result of usual non-degenerate perturbation theory expansion. However, if there exists another level with the same or close energy (this is $|m-n, N-m+n\rangle$ in our case), the self-energy itself acquires a pole in the vicinity of $\epsilon_n^{(0)}$. 
The lowest-order perturbation theory term with a pole has the order $2|m-2n|$ in $g$ because the corresponding diagram must contain at least one $G_{m-n,m-n}$ Green's function. As the perturbation operator changes the number of quanta by one, at least $2|m-2n|$ perturbation vertices are required.

Because of the pole in the self-energy term in the vicinity of $\epsilon_n^{(0)}$,
the Green's function poles positions cannot be found as simple perturbation series. For this case, it is convenient to consider the matrix Green's function 
\begin{equation} \label{honest_Green}
\mathcal{G}=\left[\begin{array}{ll}
G_{n, n} & G_{n, m-n} \\
G_{m-n, n} & G_{m-n, m-n}
\end{array}\right] .
\end{equation}

In terms of matrix Green's function, Dyson equation could also be written with self-energy matrix $\Sigma$
\begin{equation} \label{self_energ}
\begin{gathered}
\mathcal{G}=\mathcal{G}^{(0)}+\mathcal{G}^{(0)} \Sigma \mathcal{G}, \\
\mathcal{G}^{(0)} = \left[\begin{array}{cc}
\left(\omega - \epsilon_n^{(0)}\right)^{-1} & 0 \\
0 & \left(\omega - \epsilon_{m-n}^{(0)}\right)^{-1}
\end{array}\right] , \quad
\Sigma=\left[\begin{array}{cc}
\Sigma_{n, n}(\omega) & \Sigma_{n, m-n}(\omega) \\
\Sigma_{m-n, n}(\omega) & \Sigma_{m-n, m-n}(\omega)
\end{array}\right] .
\end{gathered}
\end{equation}
The solution of Eq.~\eqref{self_energ} reads
\begin{equation} \label{solut_Green}
\begin{gathered}
\mathcal{G} = \left[ \left( \mathcal{G}^{(0)} \right)^{-1} - \Sigma  \right]^{-1} = 
 \begin{bmatrix}
\omega - \epsilon_n^{(0)} - \Sigma_{n,n}(\omega) & -\Sigma_{n,m-n}(\omega) \\
-\Sigma_{m-n,n}(\omega) & \omega - \epsilon_{m-n}^{(0)} - \Sigma_{m-n,m-n}(\omega)
\end{bmatrix}^{-1}.
\end{gathered}
\end{equation}
The oscillator eigenenergies are the poles of the Green's function. They could be found from the equation $\det \left(  \mathcal{G}^{-1} \right) = 0$,
and all their dependence on the perturbation is contained in the self-energy matrix $\Sigma$. 
The diagonal terms $\Sigma_{n,n}$ and $\Sigma_{m-n,m-n}$  correspond to non-resonant energy shifts and contain contributions of all even powers of $g$.
In contrast, the off-diagonal term $\Sigma_{n,m-n}$ is proportional to $g^{m-2n}$ and is responsible for resonant multi-photon transition between the 
states $\ket{n, N-n}$ and $\ket{m-n, N-m+n}$. The leading term of its expansion in powers of $g$ reads (see~Appendix~\ref{E})
\begin{equation}\label{leading_term}
\Sigma_{n, m-n}(\omega)=g^{m-2 n} \sqrt{\frac{(m-n) !}{n !} \frac{(N-n) !}{(N-(m-n)) !}} 
\frac{1}{\left(\omega-\epsilon^{(0)}_{n+1}\right) \ldots\left(\omega-\epsilon^{(0)}_{m-n-1}\right)}+O\left(g^{m-2 n + 1}\right).
\end{equation}

Eqs.~\eqref{self_energ}-\eqref{leading_term} explain the possibility to use the non-degenerate perturbation theory up to the $|m-2n|$-th order. According to Eq.~\eqref{leading_term}, the off-diagonal terms in \eqref{solut_Green} (which correspond to the multi-photon resonance), 
have the order of $g^{m-2n}$. Therefore, they do not contribute to the $k$-th order of the perturbation theory when $k < |m-2n|$.
For $k < |m-2n|$, the energies obtained from the secular equation will coincide with those obtained from the non-degenerate 
perturbation theory. To account for multi-photon transitions, one should consider the perturbation theory of order $k \geqslant |m-2n|$.


The term $\Sigma_{n, m-n}\left(\omega = \epsilon_n^{(0)}\right)$ can be interpreted as 
the multi-photon transition amplitude between $\ket{n, N-n}$ and $\ket{m-n, N-m+n}$, and $\left| \Sigma_{n, m-n}\left(\epsilon_n^{(0)}\right) \right|$ 
equals the frequency of the multi-photon Rabi transitions $\omega^R_{n,m-n}$. 
Further, we will demonstrate that it can be treated as the tunneling amplitude in the quasiclassical limit.

\section{Classical limit}
\label{sec:classical_limit}
Multi-photon transitions described above with help of the formalism of the
perturbation theory also have a quasi-classical interpretation as tunnelling
transitions. For that, one should consider the classical limit of the 
studied model which is valid for the large number of quanta: $N \gg 1$, $\mu_N \gg 1$, 
$N/\mu_N = \mathrm{const}$.
To obtain the classical Hamiltonian of the system, 
one should replace the ladder operators in \eqref{ham} with classical 
complex amplitudes \cite{carm, mankododonov}. This results in the following complex 
Hamilton function

\begin{equation}\label{clas}
H = \omega_1 |a|^2 + \omega_2 |b|^2 + 
\frac{\alpha_1}{2}|a|^4 + \frac{\alpha_2}{2} |b|^4 + g(a^*b + b^* a).
\end{equation} 
Due to the conservation of the total number of quanta $N = |a|^2 + |b|^2$, the 
classical dynamics governed by this Hamiltonian can be described as the 
dynamics on the surface of the two-dimensional sphere. To show that, one should rewrite the 
classical Hamiltonian with help of the new 
pseudo angular momentum variables $L_x$, $L_y$, $L_z$ defined as 
\begin{equation} \label{class_ang}
\begin{gathered}
L_z=\frac{1}{2}\left(|a|^2  - |b|^2\right), \quad
  L_{+}=a^*b,  \quad L_{-}= ab^*, \\
  L_x = \frac{1}{2}\left(L_{+} + L_{-}\right), \quad L_y = \frac{1}{2i}\left( L_{+} - L_{-}\right).
\end{gathered}
\end{equation}

\begin{figure}[h!]
\center{\includegraphics[width=\textwidth]{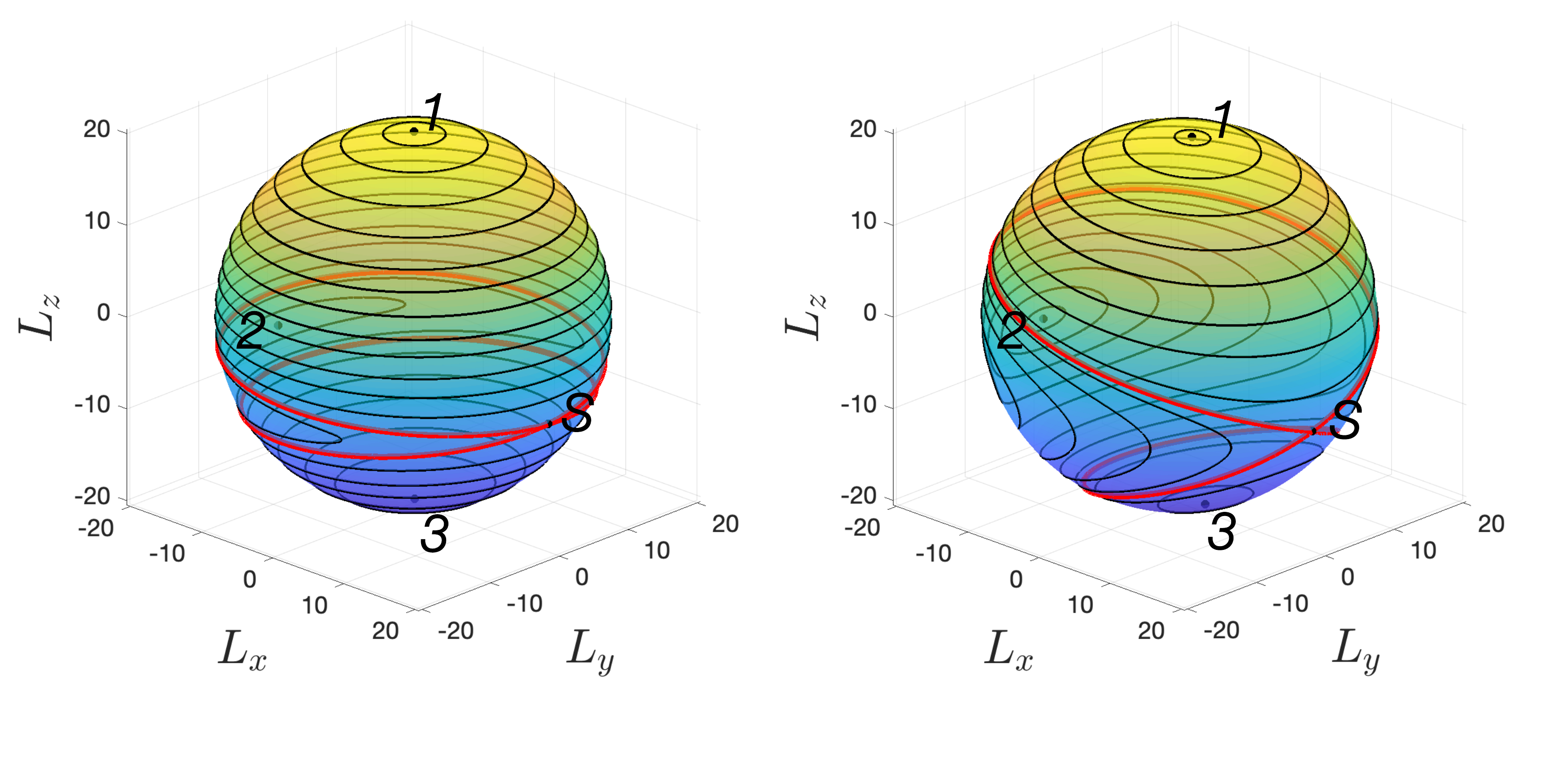}}
\caption{Classical trajectories on the surface of Bloch sphere. Black dots correspond to equilibrium points (three stable -- 1, 2, 3, and one unstable -- $S$). The red curve is a Separatrix. Here the parameters are as follows: $N=40$, $\alpha_2/\alpha_1 = 0.5$, $\Delta/\alpha_1 = 0.25$, $\mu_N = 27$, $g/\alpha_1 \approx 0.1211$ (left), and $g/\alpha_1 \approx 1.816$ (right).
This corresponds to the dimensionless coupling strength: $\sqrt{\beta} \approx 0.0103$ (left), and $\sqrt{\beta} \approx 0.1544$ (right).}
\label{fig:classical_sphere}
\end{figure}

In terms of the components of $\vec{L}$, the classical Hamiltonian
\eqref{clas} takes the form
\begin{equation} \label{ham_cons}
H = -\Delta \left( L_z + \frac{N}{2}\right) + \frac{\alpha_1}{2}\left( L_z + \frac{N}{2} \right)^2 +  
\frac{\alpha_2}{2}\left( L_z - \frac{N}{2} \right)^2  + 2gL_x + \omega_2 N.
\end{equation}
The conservation of the total angular momentum 
$L^{2}=L_{x}^{2}+L_{y}^{2}+L_{z}^{2} = N(N+2)/4$
(which is equivalent to the conservation of the total number of quanta 
in the oscillators) allows 
describing the classical dynamics with help of the classical phase portrait 
on the Bloch sphere (see Fig.~\ref{fig:classical_sphere}): 
the trajectories in the $L_{i}$ space are defined by
the conservation of the number of quanta and the 
Hamilton function \eqref{ham_cons}.

\begin{figure}[h!]
\center{\includegraphics[width=\textwidth]{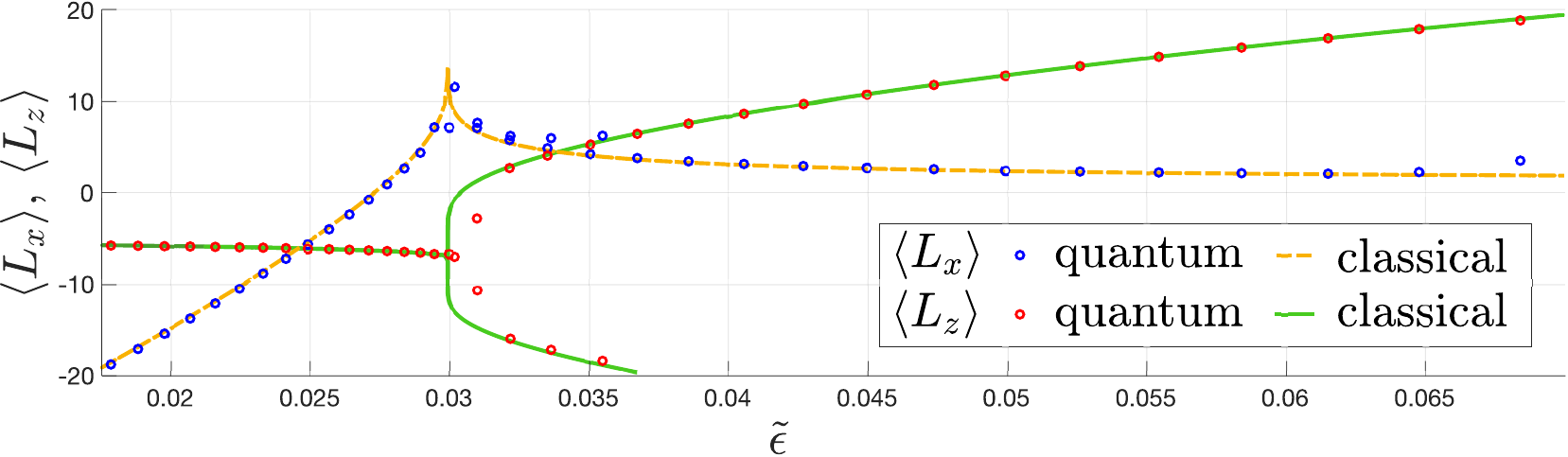}}
\caption{The average values of the pseudo angular momenta components $L_x$ and $L_z$ over the classical trajectories with the dimensionless 
energy $\tilde\epsilon$ (orange dashed and green solid line) and over the eigenstates of the quantum Hamiltonian with the eigenenergies 
$\tilde\epsilon$ (red and blue circles).
The averages are calculated at the following values of the Hamiltonian parameters: 
 $N=40$, $\alpha_2/\alpha_1 = 0.5$, $\Delta/\alpha_1 = 0.25$, $\mu_N = 27$, and $g/\alpha_1 \approx 1.816$.}
\label{fig:angular_momenta}
\end{figure}

Due to the quantum-classical correspondence, the eigenstates of the quantum Hamiltonian \eqref{ham}
correspond to a discrete set of classical trajectories on the Bloch 
sphere. We demonstrate this by comparing the period-averaged values 
of the classical momenta $L_x$, $L_z$ with the quantum-mechanical averages over the eigenstates of \eqref{ham} for non-integer $\mu_N$
in Fig.~\ref{fig:angular_momenta} (the average of $L_y$ equals zero).
As one can see from Fig.~\ref{fig:angular_momenta}, the averages calculated from the classical model are close to the quantum averages everywhere
except the vicinity of the separatrix.

Let us discuss the structure of the phase portrait in detail.
At $g = 0$, the Hamiltonian is a function of $L_z$ only, therefore,
the classical trajectories are the circles in the $L_x$, $L_y$ plane.
At non-zero $g$, the trajectories are no longer concentric circles. 
Also, a self-intersecting trajectory (separatrix) emerges which divides
the phase portrait into three regions with a stable point
inside each one (denoted as <<1>>, <<2>> and <<3>>, see~Appendix~\ref{B}). At a certain 
critical value of $g = g_\mathrm{crit}$, a saddle-node bifurcation occurs,
and one of the stable point merges with the unstable stationary 
state <<$S$>>. At larger $g$, only two stable points remain.
Depending on the value of the ratio $N/\mu_N$, the unstable point 
<<$S$>> can merge with the point <<1>> or with the point <<3>>.
If $N/\mu_N > 1$, the point <<3>> merges with the point <<$S$>>, 
otherwise, the point <<1>> merges with point <<$S$>>.

The value of $g_\mathrm{crit}$ can be found by analysing the positions of the extrema of the Hamiltonian \eqref{ham_cons} as a function of $L_i$. 
After casting the coupling constant to a dimensionless form $\beta = {8g^2N}/{(\alpha_1 + \alpha_2)^2\mu_N^3}$ (the dimensionless coupling strength), 
we obtained the following expression for $\beta_\mathrm{crit}$ (see derivation in~Supplementary~Information):
\begin{equation}
  \label{eq:beta_crit}
\beta_\mathrm{crit} = \frac{\gamma}{2}\left(\gamma^{\,2/3} - (\gamma-1)^{ 2/3}\right)^3,
\end{equation}
where $\gamma = N/\mu_N$.
Also, in the limit of $\gamma\to\infty$, one recovers the result for the model of a single Kerr oscillator with classical driving:
at $\gamma \to \infty$, $\beta_\text{crit} \to 4/27$ \cite{vogelriskenbistable, maslovagippiuskerr}.
    
\begin{figure}[h]
\center{\includegraphics[width=0.75\textwidth]{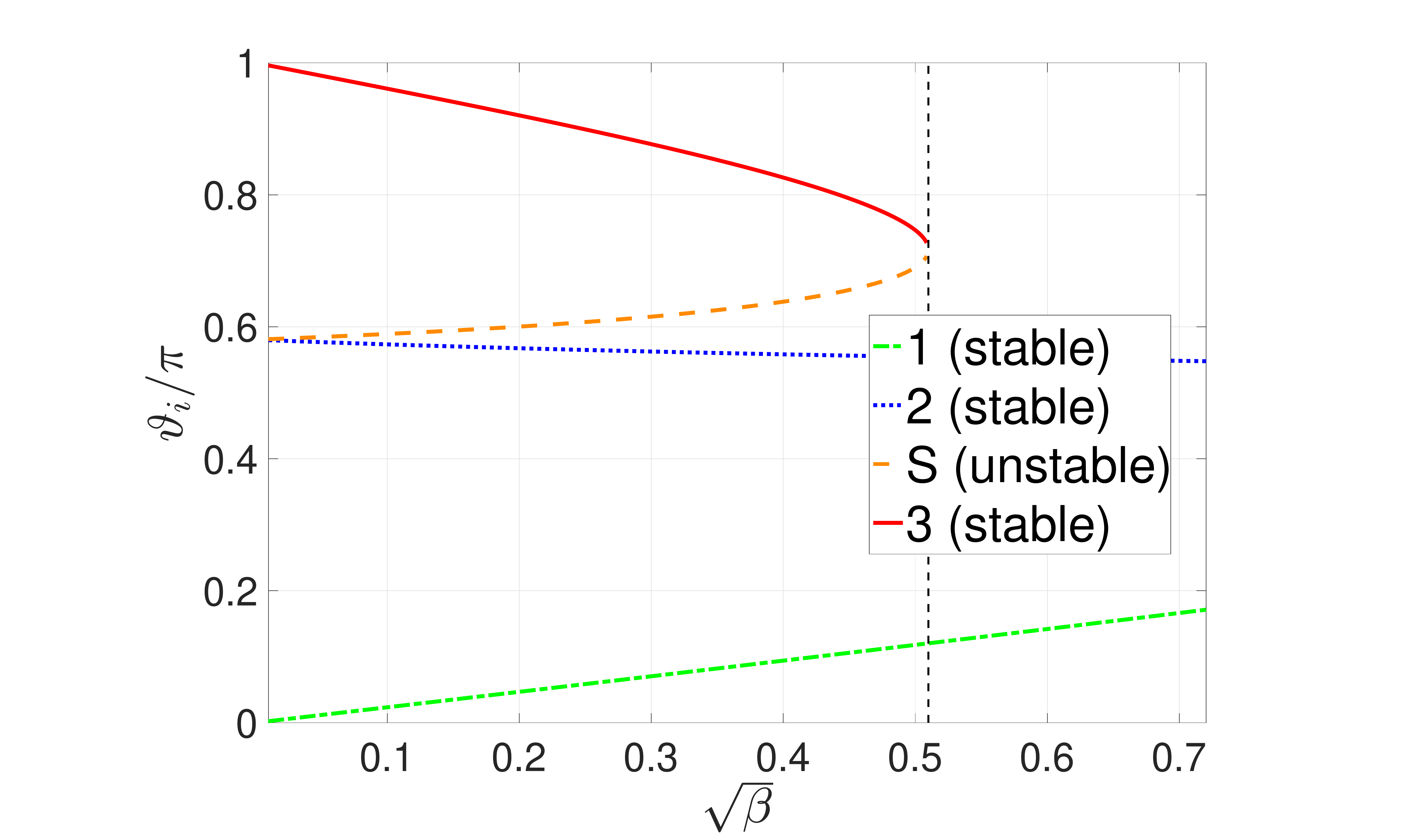}}
\caption{
The angles $\vartheta_i = \pi/2 -  \arctan(L_z^{(i)}/\big|L_x^{(i)}\big|)$ as a function of $\sqrt{\beta}$ for the $i$-th equilibrium point, 
$i  \in \{1,2,3,S \}$. Black dashed line corresponds to $\beta = \beta_\text{crit}$. The 
points numbers correspond to the ones in Fig.~\ref{fig:classical_sphere}. 
Here the parameters are as follows: $N=40$, $\alpha_2 / \alpha_1 = 0.5$, $\Delta/ \alpha_1 = 2.5$, $\mu_N = 30$, and $g = \sqrt{\beta (\Delta+\alpha_2 N)^3/N(\alpha_1 + \alpha_2)}$. 
}
\label{fig:classical}
\end{figure}

The dependence of the positions of the stable states on the dimensionless coupling strength $\beta$ is presented in Fig.~\ref{fig:classical}. 
For each of the equilibrium points $i \in \{1,2,3,S \}$, the value of the polar angle $\vartheta_i = \pi/2 -  \arctan(L_z^{(i)}/\big|L_x^{(i)}\big|)$ is 
plotted (see~Supplementary~Information for more details). At $\beta = \beta_\text{crit}$ (vertical black dotted line),
the states <<$S$>> (unstable) and <<3>> (stable) merge. 
Also, both of the angles $\vartheta_1$ and $\vartheta_2$ approach $\pi/2$ at $\beta\gg\beta_\text{crit}$, 
which corresponds to the diametrically opposite stationary points in the $X$-$Y$ plane.

Now let us discuss how tunnelling modifies the classical picture 
and establish the relation between tunnelling and multi-photon transitions. 
Tunnelling transitions are possible providing that there exist classical trajectories with the same energies. Because of the structure
of the classical phase portrait, this holds for $g < g_\mathrm{crit}$.
In this case, tunnelling transitions are possible between the classical trajectories from regions 1 and 3.  
At small $g$, the classical trajectories are close to the circles in the $X$-$Y$ plane (see Fig.~\ref{fig:classical_sphere}, left panel) 
and can be identified with the oscillators Fock states: the Fock state $|n,N-n\rangle$ corresponds to the circle with $L^2 = N^2/4$, $L_z = N/2 - n$. For this case, one can directly apply the results of Sections \ref{sec:the_model} and 
\ref{sec:symmetry_proof} 
and deduce that at the integer $\mu_N$, resonant tunnelling transitions are possible between the trajectories with $L_z = N/2 - n$ 
and $L_z = N/2 - \mu_N + n$. For larger coupling values, the eigenstates are no longer the Fock states, and the 
classical trajectories are no longer circles in the $X$-$Y$ plane. 
However, we state that the condition for resonant tunnelling remains the same for all values of $g \in [0, g_\mathrm{crit}]$. 
This is supported by the fact that the calculation of section \ref{sec:symmetry_proof} is performed in all orders of the perturbation theory. 
Because of that, it fully takes into account the modification of the classical trajectories at larger $g$. In addition, the numerical diagonalization 
of the Hamiltonian \eqref{ham} demonstrates that the behaviour shown in Fig.~\ref{fig:spectra} (simultaneous anticrossings at the integer $\mu_N$) persists for the whole range $g \in [0, g_\mathrm{crit}]$.

\section*{Conclusions}
For the model of two coupled quantum oscillators with Kerr nonlinearities and linear coupling, we studied the multi-photon 
transitions between the oscillators. We showed that for certain parameters of the model, the resonant condition for multi-photon transitions is simultaneously satisfied for many pairs of the oscillator Fock states. This holds even for the moderate 
coupling strength between oscillators, and this is the consequence of a special symmetry of the oscillators Hamiltonian. 

The latter is related to the structure of the perturbation series of 
the model eigenenergies and was proven with help of analytical continuation
of the Hamiltonian for non-integer numbers of quanta.

Also, in the quasi-classical limit,
the phase space of the two coupled oscillators can be mapped on a sphere, and the multi-photon transitions can be interpreted as tunnelling 
transitions between the trajectories lying in different regions of the classical phase space. 
Thus, when the resonant condition is satisfied, tunnel transitions affect the whole region of the classical phase space. 

We believe that the results obtained in this work could be relevant for the experiments involving
high-quality oscillator modes with low occupation numbers, such as the plasmon modes of the Josephson junction arrays or phonon modes of trapped ions ensembles. In particular, the independence of the multi-photon resonances positions could be used for the measurements of the Kerr coefficients of the modes. Also, the multi-photon transitions provide a way to create the entangled states of two oscillators. In addition, the obtained results could be used for certain models of dissipative time crystals in the quantum regime, and the symmetry discovered in the considered model may allow obtaining new exact results in quantum-optical systems.

\begin{acknowledgements}
This work was supported by the RFBR grants 19--02--000--87a, 18--29--20032mk, 19-32-90169, 
and by a grant of the Foundation for the Advancement of Theoretical 
Physics and Mathematics 'Basis'.
Natalya S. Maslova thanks Russian Science Foundation grant 
(Project 18--72--10002) for the support of the work on the results presented
in Section~\ref{sec:high_ord}.

\end{acknowledgements}
\bibliography{sample}

\begin{thebibliography}{37}%
\makeatletter
\providecommand \@ifxundefined [1]{%
 \@ifx{#1\undefined}
}%
\providecommand \@ifnum [1]{%
 \ifnum #1\expandafter \@firstoftwo
 \else \expandafter \@secondoftwo
 \fi
}%
\providecommand \@ifx [1]{%
 \ifx #1\expandafter \@firstoftwo
 \else \expandafter \@secondoftwo
 \fi
}%
\providecommand \natexlab [1]{#1}%
\providecommand \enquote  [1]{``#1''}%
\providecommand \bibnamefont  [1]{#1}%
\providecommand \bibfnamefont [1]{#1}%
\providecommand \citenamefont [1]{#1}%
\providecommand \href@noop [0]{\@secondoftwo}%
\providecommand \href [0]{\begingroup \@sanitize@url \@href}%
\providecommand \@href[1]{\@@startlink{#1}\@@href}%
\providecommand \@@href[1]{\endgroup#1\@@endlink}%
\providecommand \@sanitize@url [0]{\catcode `\\12\catcode `\$12\catcode
  `\&12\catcode `\#12\catcode `\^12\catcode `\_12\catcode `\%12\relax}%
\providecommand \@@startlink[1]{}%
\providecommand \@@endlink[0]{}%
\providecommand \url  [0]{\begingroup\@sanitize@url \@url }%
\providecommand \@url [1]{\endgroup\@href {#1}{\urlprefix }}%
\providecommand \urlprefix  [0]{URL }%
\providecommand \Eprint [0]{\href }%
\providecommand \doibase [0]{https://doi.org/}%
\providecommand \selectlanguage [0]{\@gobble}%
\providecommand \bibinfo  [0]{\@secondoftwo}%
\providecommand \bibfield  [0]{\@secondoftwo}%
\providecommand \translation [1]{[#1]}%
\providecommand \BibitemOpen [0]{}%
\providecommand \bibitemStop [0]{}%
\providecommand \bibitemNoStop [0]{.\EOS\space}%
\providecommand \EOS [0]{\spacefactor3000\relax}%
\providecommand \BibitemShut  [1]{\csname bibitem#1\endcsname}%
\let\auto@bib@innerbib\@empty
\bibitem [{\citenamefont {Adamyan}\ \emph {et~al.}(2001)\citenamefont
  {Adamyan}, \citenamefont {Manvelyan},\ and\ \citenamefont
  {Kryuchkyan}}]{Adamyan2001}%
  \BibitemOpen
  \bibfield  {author} {\bibinfo {author} {\bibfnamefont {H.~H.}\ \bibnamefont
  {Adamyan}}, \bibinfo {author} {\bibfnamefont {S.~B.}\ \bibnamefont
  {Manvelyan}},\ and\ \bibinfo {author} {\bibfnamefont {G.~Y.}\ \bibnamefont
  {Kryuchkyan}},\ }\bibfield  {title} {\bibinfo {title} {Chaos in a double
  driven dissipative nonlinear oscillator},\ }\href
  {https://doi.org/10.1103/physreve.64.046219} {\bibfield  {journal} {\bibinfo
  {journal} {Physical Review E}\ }\textbf {\bibinfo {volume} {64}},\ \bibinfo
  {pages} {046219} (\bibinfo {year} {2001})}\BibitemShut {NoStop}%
\bibitem [{\citenamefont {Tadokoro}\ \emph {et~al.}(2018)\citenamefont
  {Tadokoro}, \citenamefont {Tanaka},\ and\ \citenamefont
  {Dykman}}]{Tadokoro2018}%
  \BibitemOpen
  \bibfield  {author} {\bibinfo {author} {\bibfnamefont {Y.}~\bibnamefont
  {Tadokoro}}, \bibinfo {author} {\bibfnamefont {H.}~\bibnamefont {Tanaka}},\
  and\ \bibinfo {author} {\bibfnamefont {M.~I.}\ \bibnamefont {Dykman}},\
  }\bibfield  {title} {\bibinfo {title} {Driven nonlinear nanomechanical
  resonators as digital signal detectors},\ }\bibfield  {journal} {\bibinfo
  {journal} {Scientific Reports}\ }\textbf {\bibinfo {volume} {8}},\ \href
  {https://doi.org/10.1038/s41598-018-29572-7} {10.1038/s41598-018-29572-7}
  (\bibinfo {year} {2018})\BibitemShut {NoStop}%
\bibitem [{\citenamefont {Zhang}\ and\ \citenamefont
  {Baranger}(2021)}]{Zhang2021}%
  \BibitemOpen
  \bibfield  {author} {\bibinfo {author} {\bibfnamefont {X.~H.~H.}\
  \bibnamefont {Zhang}}\ and\ \bibinfo {author} {\bibfnamefont {H.~U.}\
  \bibnamefont {Baranger}},\ }\bibfield  {title} {\bibinfo {title}
  {Driven-dissipative phase transition in a kerr oscillator: From semiclassical
  $\mathcal{PT}$ symmetry to quantum fluctuations},\ }\href
  {https://doi.org/10.1103/PhysRevA.103.033711} {\bibfield  {journal} {\bibinfo
   {journal} {Phys. Rev. A}\ }\textbf {\bibinfo {volume} {103}},\ \bibinfo
  {pages} {033711} (\bibinfo {year} {2021})}\BibitemShut {NoStop}%
\bibitem [{\citenamefont {Serban}\ and\ \citenamefont
  {Wilhelm}(2007)}]{Serban2007}%
  \BibitemOpen
  \bibfield  {author} {\bibinfo {author} {\bibfnamefont {I.}~\bibnamefont
  {Serban}}\ and\ \bibinfo {author} {\bibfnamefont {F.~K.}\ \bibnamefont
  {Wilhelm}},\ }\bibfield  {title} {\bibinfo {title} {Dynamical tunneling in
  macroscopic systems},\ }\href {https://doi.org/10.1103/PhysRevLett.99.137001}
  {\bibfield  {journal} {\bibinfo  {journal} {Phys. Rev. Lett.}\ }\textbf
  {\bibinfo {volume} {99}},\ \bibinfo {pages} {137001} (\bibinfo {year}
  {2007})}\BibitemShut {NoStop}%
\bibitem [{\citenamefont {Goto}(2016)}]{Goto2016}%
  \BibitemOpen
  \bibfield  {author} {\bibinfo {author} {\bibfnamefont {H.}~\bibnamefont
  {Goto}},\ }\bibfield  {title} {\bibinfo {title} {Universal quantum
  computation with a nonlinear oscillator network},\ }\href
  {https://doi.org/10.1103/physreva.93.050301} {\bibfield  {journal} {\bibinfo
  {journal} {Physical Review A}\ }\textbf {\bibinfo {volume} {93}},\ \bibinfo
  {pages} {050301} (\bibinfo {year} {2016})}\BibitemShut {NoStop}%
\bibitem [{\citenamefont {Maslova}\ \emph
  {et~al.}(2019{\natexlab{a}})\citenamefont {Maslova}, \citenamefont
  {Mantsevich}, \citenamefont {Arseyev},\ and\ \citenamefont
  {Sokolov}}]{PhysRevB.100.035307}%
  \BibitemOpen
  \bibfield  {author} {\bibinfo {author} {\bibfnamefont {N.~S.}\ \bibnamefont
  {Maslova}}, \bibinfo {author} {\bibfnamefont {V.~N.}\ \bibnamefont
  {Mantsevich}}, \bibinfo {author} {\bibfnamefont {P.~I.}\ \bibnamefont
  {Arseyev}},\ and\ \bibinfo {author} {\bibfnamefont {I.~M.}\ \bibnamefont
  {Sokolov}},\ }\bibfield  {title} {\bibinfo {title} {Tunneling current induced
  squeezing of the single-molecule vibrational mode},\ }\href
  {https://doi.org/10.1103/PhysRevB.100.035307} {\bibfield  {journal} {\bibinfo
   {journal} {Phys. Rev. B}\ }\textbf {\bibinfo {volume} {100}},\ \bibinfo
  {pages} {035307} (\bibinfo {year} {2019}{\natexlab{a}})}\BibitemShut
  {NoStop}%
\bibitem [{\citenamefont {Joshi}\ \emph {et~al.}(2011)\citenamefont {Joshi},
  \citenamefont {Jonson}, \citenamefont {Andersson},\ and\ \citenamefont
  {Öhberg}}]{Entangled}%
  \BibitemOpen
  \bibfield  {author} {\bibinfo {author} {\bibfnamefont {C.}~\bibnamefont
  {Joshi}}, \bibinfo {author} {\bibfnamefont {M.}~\bibnamefont {Jonson}},
  \bibinfo {author} {\bibfnamefont {E.}~\bibnamefont {Andersson}},\ and\
  \bibinfo {author} {\bibfnamefont {P.}~\bibnamefont {Öhberg}},\ }\bibfield
  {title} {\bibinfo {title} {Quantum entanglement of anharmonic oscillators},\
  }\href {https://doi.org/10.1088/0953-4075/44/24/245503} {\bibfield  {journal}
  {\bibinfo  {journal} {Journal of Physics B: Atomic, Molecular and Optical
  Physics}\ }\textbf {\bibinfo {volume} {44}},\ \bibinfo {pages} {245503}
  (\bibinfo {year} {2011})}\BibitemShut {NoStop}%
\bibitem [{\citenamefont {Teh}\ \emph {et~al.}(2020)\citenamefont {Teh},
  \citenamefont {Sun}, \citenamefont {Polkinghorne}, \citenamefont {He},
  \citenamefont {Gong}, \citenamefont {Drummond},\ and\ \citenamefont
  {Reid}}]{Entangled3}%
  \BibitemOpen
  \bibfield  {author} {\bibinfo {author} {\bibfnamefont {R.~Y.}\ \bibnamefont
  {Teh}}, \bibinfo {author} {\bibfnamefont {F.-X.}\ \bibnamefont {Sun}},
  \bibinfo {author} {\bibfnamefont {R.~E.~S.}\ \bibnamefont {Polkinghorne}},
  \bibinfo {author} {\bibfnamefont {Q.~Y.}\ \bibnamefont {He}}, \bibinfo
  {author} {\bibfnamefont {Q.}~\bibnamefont {Gong}}, \bibinfo {author}
  {\bibfnamefont {P.~D.}\ \bibnamefont {Drummond}},\ and\ \bibinfo {author}
  {\bibfnamefont {M.~D.}\ \bibnamefont {Reid}},\ }\bibfield  {title} {\bibinfo
  {title} {Dynamics of transient cat states in degenerate parametric
  oscillation with and without nonlinear kerr interactions},\ }\href
  {https://doi.org/10.1103/PhysRevA.101.043807} {\bibfield  {journal} {\bibinfo
   {journal} {Phys. Rev. A}\ }\textbf {\bibinfo {volume} {101}},\ \bibinfo
  {pages} {043807} (\bibinfo {year} {2020})}\BibitemShut {NoStop}%
\bibitem [{\citenamefont {Dodonov}\ \emph {et~al.}(1974)\citenamefont
  {Dodonov}, \citenamefont {Malkin},\ and\ \citenamefont
  {Man{'}ko}}]{DODONOV1974597}%
  \BibitemOpen
  \bibfield  {author} {\bibinfo {author} {\bibfnamefont {V.}~\bibnamefont
  {Dodonov}}, \bibinfo {author} {\bibfnamefont {I.}~\bibnamefont {Malkin}},\
  and\ \bibinfo {author} {\bibfnamefont {V.}~\bibnamefont {Man{'}ko}},\
  }\bibfield  {title} {\bibinfo {title} {Even and odd coherent states and
  excitations of a singular oscillator},\ }\href
  {https://doi.org/10.1016/0031-8914(74)90215-8} {\bibfield  {journal}
  {\bibinfo  {journal} {Physica}\ }\textbf {\bibinfo {volume} {72}},\ \bibinfo
  {pages} {597} (\bibinfo {year} {1974})}\BibitemShut {NoStop}%
\bibitem [{\citenamefont {Andersen}\ \emph {et~al.}(2020)\citenamefont
  {Andersen}, \citenamefont {Kamal}, \citenamefont {Masluk}, \citenamefont
  {Pop}, \citenamefont {Blais},\ and\ \citenamefont {Devoret}}]{Andersen2020}%
  \BibitemOpen
  \bibfield  {author} {\bibinfo {author} {\bibfnamefont {C.~K.}\ \bibnamefont
  {Andersen}}, \bibinfo {author} {\bibfnamefont {A.}~\bibnamefont {Kamal}},
  \bibinfo {author} {\bibfnamefont {N.~A.}\ \bibnamefont {Masluk}}, \bibinfo
  {author} {\bibfnamefont {I.~M.}\ \bibnamefont {Pop}}, \bibinfo {author}
  {\bibfnamefont {A.}~\bibnamefont {Blais}},\ and\ \bibinfo {author}
  {\bibfnamefont {M.~H.}\ \bibnamefont {Devoret}},\ }\bibfield  {title}
  {\bibinfo {title} {Quantum versus classical switching dynamics of driven
  dissipative kerr resonators},\ }\href
  {https://doi.org/10.1103/PhysRevApplied.13.044017} {\bibfield  {journal}
  {\bibinfo  {journal} {Phys. Rev. Appl.}\ }\textbf {\bibinfo {volume} {13}},\
  \bibinfo {pages} {044017} (\bibinfo {year} {2020})}\BibitemShut {NoStop}%
\bibitem [{\citenamefont {Pistolesi}(2018)}]{Pistolesi2018}%
  \BibitemOpen
  \bibfield  {author} {\bibinfo {author} {\bibfnamefont {F.}~\bibnamefont
  {Pistolesi}},\ }\bibfield  {title} {\bibinfo {title} {Bistability of a slow
  mechanical oscillator coupled to a laser-driven two-level system},\ }\href
  {https://doi.org/10.1103/PhysRevA.97.063833} {\bibfield  {journal} {\bibinfo
  {journal} {Phys. Rev. A}\ }\textbf {\bibinfo {volume} {97}},\ \bibinfo
  {pages} {063833} (\bibinfo {year} {2018})}\BibitemShut {NoStop}%
\bibitem [{\citenamefont {Ding}\ \emph {et~al.}(2017)\citenamefont {Ding},
  \citenamefont {Maslennikov}, \citenamefont {Hablützel}, \citenamefont
  {Loh},\ and\ \citenamefont {Matsukevich}}]{Ding2017}%
  \BibitemOpen
  \bibfield  {author} {\bibinfo {author} {\bibfnamefont {S.}~\bibnamefont
  {Ding}}, \bibinfo {author} {\bibfnamefont {G.}~\bibnamefont {Maslennikov}},
  \bibinfo {author} {\bibfnamefont {R.}~\bibnamefont {Hablützel}}, \bibinfo
  {author} {\bibfnamefont {H.}~\bibnamefont {Loh}},\ and\ \bibinfo {author}
  {\bibfnamefont {D.}~\bibnamefont {Matsukevich}},\ }\bibfield  {title}
  {\bibinfo {title} {Quantum parametric oscillator with trapped ions},\ }\href
  {https://doi.org/10.1103/physrevlett.119.150404} {\bibfield  {journal}
  {\bibinfo  {journal} {Physical Review Letters}\ }\textbf {\bibinfo {volume}
  {119}},\ \bibinfo {pages} {150404} (\bibinfo {year} {2017})}\BibitemShut
  {NoStop}%
\bibitem [{\citenamefont {Muppalla}\ \emph {et~al.}(2018)\citenamefont
  {Muppalla}, \citenamefont {Gargiulo}, \citenamefont {Mirzaei}, \citenamefont
  {Venkatesh}, \citenamefont {Juan}, \citenamefont {Gr\"unhaupt}, \citenamefont
  {Pop},\ and\ \citenamefont {Kirchmair}}]{Muppalla}%
  \BibitemOpen
  \bibfield  {author} {\bibinfo {author} {\bibfnamefont {P.~R.}\ \bibnamefont
  {Muppalla}}, \bibinfo {author} {\bibfnamefont {O.}~\bibnamefont {Gargiulo}},
  \bibinfo {author} {\bibfnamefont {S.~I.}\ \bibnamefont {Mirzaei}}, \bibinfo
  {author} {\bibfnamefont {B.~P.}\ \bibnamefont {Venkatesh}}, \bibinfo {author}
  {\bibfnamefont {M.~L.}\ \bibnamefont {Juan}}, \bibinfo {author}
  {\bibfnamefont {L.}~\bibnamefont {Gr\"unhaupt}}, \bibinfo {author}
  {\bibfnamefont {I.~M.}\ \bibnamefont {Pop}},\ and\ \bibinfo {author}
  {\bibfnamefont {G.}~\bibnamefont {Kirchmair}},\ }\bibfield  {title} {\bibinfo
  {title} {Bistability in a mesoscopic josephson junction array resonator},\
  }\href {https://doi.org/10.1103/PhysRevB.97.024518} {\bibfield  {journal}
  {\bibinfo  {journal} {Phys. Rev. B}\ }\textbf {\bibinfo {volume} {97}},\
  \bibinfo {pages} {024518} (\bibinfo {year} {2018})}\BibitemShut {NoStop}%
\bibitem [{\citenamefont {Siddiqi}\ \emph {et~al.}(2005)\citenamefont
  {Siddiqi}, \citenamefont {Vijay}, \citenamefont {Pierre}, \citenamefont
  {Wilson}, \citenamefont {Frunzio}, \citenamefont {Metcalfe}, \citenamefont
  {Rigetti}, \citenamefont {Schoelkopf}, \citenamefont {Devoret}, \citenamefont
  {Vion},\ and\ \citenamefont {Esteve}}]{Siddiqi2005}%
  \BibitemOpen
  \bibfield  {author} {\bibinfo {author} {\bibfnamefont {I.}~\bibnamefont
  {Siddiqi}}, \bibinfo {author} {\bibfnamefont {R.}~\bibnamefont {Vijay}},
  \bibinfo {author} {\bibfnamefont {F.}~\bibnamefont {Pierre}}, \bibinfo
  {author} {\bibfnamefont {C.~M.}\ \bibnamefont {Wilson}}, \bibinfo {author}
  {\bibfnamefont {L.}~\bibnamefont {Frunzio}}, \bibinfo {author} {\bibfnamefont
  {M.}~\bibnamefont {Metcalfe}}, \bibinfo {author} {\bibfnamefont
  {C.}~\bibnamefont {Rigetti}}, \bibinfo {author} {\bibfnamefont {R.~J.}\
  \bibnamefont {Schoelkopf}}, \bibinfo {author} {\bibfnamefont {M.~H.}\
  \bibnamefont {Devoret}}, \bibinfo {author} {\bibfnamefont {D.}~\bibnamefont
  {Vion}},\ and\ \bibinfo {author} {\bibfnamefont {D.}~\bibnamefont {Esteve}},\
  }\bibfield  {title} {\bibinfo {title} {Direct observation of dynamical
  bifurcation between two driven oscillation states of a josephson junction},\
  }\href {https://doi.org/10.1103/PhysRevLett.94.027005} {\bibfield  {journal}
  {\bibinfo  {journal} {Phys. Rev. Lett.}\ }\textbf {\bibinfo {volume} {94}},\
  \bibinfo {pages} {027005} (\bibinfo {year} {2005})}\BibitemShut {NoStop}%
\bibitem [{\citenamefont {Shirai}\ \emph {et~al.}(2018)\citenamefont {Shirai},
  \citenamefont {Todo}, \citenamefont {de~Raedt},\ and\ \citenamefont
  {Miyashita}}]{Shirai2018}%
  \BibitemOpen
  \bibfield  {author} {\bibinfo {author} {\bibfnamefont {T.}~\bibnamefont
  {Shirai}}, \bibinfo {author} {\bibfnamefont {S.}~\bibnamefont {Todo}},
  \bibinfo {author} {\bibfnamefont {H.}~\bibnamefont {de~Raedt}},\ and\
  \bibinfo {author} {\bibfnamefont {S.}~\bibnamefont {Miyashita}},\ }\bibfield
  {title} {\bibinfo {title} {Optical bistability in a low-photon-density
  regime},\ }\href {https://doi.org/10.1103/PhysRevA.98.043802} {\bibfield
  {journal} {\bibinfo  {journal} {Phys. Rev. A}\ }\textbf {\bibinfo {volume}
  {98}},\ \bibinfo {pages} {043802} (\bibinfo {year} {2018})}\BibitemShut
  {NoStop}%
\bibitem [{\citenamefont {Maslova}\ \emph
  {et~al.}(2019{\natexlab{b}})\citenamefont {Maslova}, \citenamefont {Anikin},
  \citenamefont {Mantsevich}, \citenamefont {Gippius},\ and\ \citenamefont
  {Sokolov}}]{11}%
  \BibitemOpen
  \bibfield  {author} {\bibinfo {author} {\bibfnamefont {N.~S.}\ \bibnamefont
  {Maslova}}, \bibinfo {author} {\bibfnamefont {E.~V.}\ \bibnamefont {Anikin}},
  \bibinfo {author} {\bibfnamefont {V.~N.}\ \bibnamefont {Mantsevich}},
  \bibinfo {author} {\bibfnamefont {N.~A.}\ \bibnamefont {Gippius}},\ and\
  \bibinfo {author} {\bibfnamefont {I.~M.}\ \bibnamefont {Sokolov}},\
  }\bibfield  {title} {\bibinfo {title} {Quantum tunneling effect on switching
  rates of bistable driven system},\ }\href
  {https://doi.org/10.1088/1612-202x/ab0a59} {\bibfield  {journal} {\bibinfo
  {journal} {Laser Physics Letters}\ }\textbf {\bibinfo {volume} {16}},\
  \bibinfo {pages} {045205} (\bibinfo {year} {2019}{\natexlab{b}})}\BibitemShut
  {NoStop}%
\bibitem [{\citenamefont {Wallraff}\ \emph {et~al.}(2003)\citenamefont
  {Wallraff}, \citenamefont {Duty}, \citenamefont {Lukashenko},\ and\
  \citenamefont {Ustinov}}]{tunMP1}%
  \BibitemOpen
  \bibfield  {author} {\bibinfo {author} {\bibfnamefont {A.}~\bibnamefont
  {Wallraff}}, \bibinfo {author} {\bibfnamefont {T.}~\bibnamefont {Duty}},
  \bibinfo {author} {\bibfnamefont {A.}~\bibnamefont {Lukashenko}},\ and\
  \bibinfo {author} {\bibfnamefont {A.~V.}\ \bibnamefont {Ustinov}},\
  }\bibfield  {title} {\bibinfo {title} {Multiphoton transitions between energy
  levels in a current-biased josephson tunnel junction},\ }\href
  {https://doi.org/10.1103/physrevlett.90.037003} {\bibfield  {journal}
  {\bibinfo  {journal} {Physical Review Letters}\ }\textbf {\bibinfo {volume}
  {90}},\ \bibinfo {pages} {037003} (\bibinfo {year} {2003})}\BibitemShut
  {NoStop}%
\bibitem [{\citenamefont {Dykman}\ and\ \citenamefont
  {Fistul}(2005)}]{Dykman2005}%
  \BibitemOpen
  \bibfield  {author} {\bibinfo {author} {\bibfnamefont {M.~I.}\ \bibnamefont
  {Dykman}}\ and\ \bibinfo {author} {\bibfnamefont {M.~V.}\ \bibnamefont
  {Fistul}},\ }\bibfield  {title} {\bibinfo {title} {Multiphoton
  antiresonance},\ }\href {https://doi.org/10.1103/physrevb.71.140508}
  {\bibfield  {journal} {\bibinfo  {journal} {Physical Review B}\ }\textbf
  {\bibinfo {volume} {71}},\ \bibinfo {pages} {140508} (\bibinfo {year}
  {2005})}\BibitemShut {NoStop}%
\bibitem [{\citenamefont {Maslova}\ \emph
  {et~al.}(2019{\natexlab{c}})\citenamefont {Maslova}, \citenamefont {Anikin},
  \citenamefont {Gippius},\ and\ \citenamefont {Sokolov}}]{Anikin2018}%
  \BibitemOpen
  \bibfield  {author} {\bibinfo {author} {\bibfnamefont {N.~S.}\ \bibnamefont
  {Maslova}}, \bibinfo {author} {\bibfnamefont {E.~V.}\ \bibnamefont {Anikin}},
  \bibinfo {author} {\bibfnamefont {N.~A.}\ \bibnamefont {Gippius}},\ and\
  \bibinfo {author} {\bibfnamefont {I.~M.}\ \bibnamefont {Sokolov}},\
  }\bibfield  {title} {\bibinfo {title} {Effects of tunneling and multiphoton
  transitions on squeezed-state generation in bistable driven systems},\ }\href
  {https://doi.org/10.1103/PhysRevA.99.043802} {\bibfield  {journal} {\bibinfo
  {journal} {Phys. Rev. A}\ }\textbf {\bibinfo {volume} {99}},\ \bibinfo
  {pages} {043802} (\bibinfo {year} {2019}{\natexlab{c}})}\BibitemShut
  {NoStop}%
\bibitem [{\citenamefont {Wang}\ \emph {et~al.}(2019)\citenamefont {Wang},
  \citenamefont {Zhang}, \citenamefont {Li}, \citenamefont {Xu}, \citenamefont
  {Cao}, \citenamefont {Zhou}, \citenamefont {Cao},\ and\ \citenamefont
  {Lu}}]{tun2MP}%
  \BibitemOpen
  \bibfield  {author} {\bibinfo {author} {\bibfnamefont {R.}~\bibnamefont
  {Wang}}, \bibinfo {author} {\bibfnamefont {Q.}~\bibnamefont {Zhang}},
  \bibinfo {author} {\bibfnamefont {D.}~\bibnamefont {Li}}, \bibinfo {author}
  {\bibfnamefont {S.}~\bibnamefont {Xu}}, \bibinfo {author} {\bibfnamefont
  {P.}~\bibnamefont {Cao}}, \bibinfo {author} {\bibfnamefont {Y.}~\bibnamefont
  {Zhou}}, \bibinfo {author} {\bibfnamefont {W.}~\bibnamefont {Cao}},\ and\
  \bibinfo {author} {\bibfnamefont {P.}~\bibnamefont {Lu}},\ }\bibfield
  {title} {\bibinfo {title} {Identification of tunneling and multiphoton
  ionization in intermediate keldysh parameter regime},\ }\href
  {https://doi.org/10.1364/oe.27.006471} {\bibfield  {journal} {\bibinfo
  {journal} {Optics Express}\ }\textbf {\bibinfo {volume} {27}},\ \bibinfo
  {pages} {6471} (\bibinfo {year} {2019})}\BibitemShut {NoStop}%
\bibitem [{\citenamefont {Carmichael}(1998)}]{carm}%
  \BibitemOpen
  \bibfield  {author} {\bibinfo {author} {\bibfnamefont {H.~J.}\ \bibnamefont
  {Carmichael}},\ }\href
  {https://doi.org/https://doi.org/10.1007/978-3-662-03875-8} {\emph {\bibinfo
  {title} {Statistical Methods in Quantum Optics 1}}}\ (\bibinfo  {publisher}
  {Springer Berlin, Heidelberg},\ \bibinfo {year} {1998})\ pp.\ \bibinfo
  {pages} {XXI, 361}\BibitemShut {NoStop}%
\bibitem [{\citenamefont {Owerre}\ and\ \citenamefont
  {Paranjape}(2015)}]{OWERRE20151}%
  \BibitemOpen
  \bibfield  {author} {\bibinfo {author} {\bibfnamefont {S.}~\bibnamefont
  {Owerre}}\ and\ \bibinfo {author} {\bibfnamefont {M.}~\bibnamefont
  {Paranjape}},\ }\bibfield  {title} {\bibinfo {title} {Macroscopic quantum
  tunneling and quantum{\textendash}classical phase transitions of the escape
  rate in large spin systems},\ }\href
  {https://doi.org/10.1016/j.physrep.2014.09.001} {\bibfield  {journal}
  {\bibinfo  {journal} {Physics Reports}\ }\textbf {\bibinfo {volume} {546}},\
  \bibinfo {pages} {1} (\bibinfo {year} {2015})}\BibitemShut {NoStop}%
\bibitem [{\citenamefont {Seibold}\ \emph {et~al.}(2020)\citenamefont
  {Seibold}, \citenamefont {Rota},\ and\ \citenamefont {Savona}}]{DTC1}%
  \BibitemOpen
  \bibfield  {author} {\bibinfo {author} {\bibfnamefont {K.}~\bibnamefont
  {Seibold}}, \bibinfo {author} {\bibfnamefont {R.}~\bibnamefont {Rota}},\ and\
  \bibinfo {author} {\bibfnamefont {V.}~\bibnamefont {Savona}},\ }\bibfield
  {title} {\bibinfo {title} {Dissipative time crystal in an asymmetric
  nonlinear photonic dimer},\ }\href
  {https://doi.org/10.1103/PhysRevA.101.033839} {\bibfield  {journal} {\bibinfo
   {journal} {Phys. Rev. A}\ }\textbf {\bibinfo {volume} {101}},\ \bibinfo
  {pages} {033839} (\bibinfo {year} {2020})}\BibitemShut {NoStop}%
\bibitem [{\citenamefont {Lled{\'{o}}}\ and\ \citenamefont
  {Szyma{\'{n}}ska}(2020)}]{DTC2}%
  \BibitemOpen
  \bibfield  {author} {\bibinfo {author} {\bibfnamefont {C.}~\bibnamefont
  {Lled{\'{o}}}}\ and\ \bibinfo {author} {\bibfnamefont {M.~H.}\ \bibnamefont
  {Szyma{\'{n}}ska}},\ }\bibfield  {title} {\bibinfo {title} {A dissipative
  time crystal with or without $\mathbb{Z}_2$ symmetry breaking},\ }\href
  {https://doi.org/10.1088/1367-2630/ab9ae3} {\bibfield  {journal} {\bibinfo
  {journal} {New Journal of Physics}\ }\textbf {\bibinfo {volume} {22}},\
  \bibinfo {pages} {075002} (\bibinfo {year} {2020})}\BibitemShut {NoStop}%
\bibitem [{\citenamefont {Sacha}\ and\ \citenamefont
  {Zakrzewski}(2017)}]{DTCrev1}%
  \BibitemOpen
  \bibfield  {author} {\bibinfo {author} {\bibfnamefont {K.}~\bibnamefont
  {Sacha}}\ and\ \bibinfo {author} {\bibfnamefont {J.}~\bibnamefont
  {Zakrzewski}},\ }\bibfield  {title} {\bibinfo {title} {Time crystals: a
  review},\ }\href {https://doi.org/10.1088/1361-6633/aa8b38} {\bibfield
  {journal} {\bibinfo  {journal} {Reports on Progress in Physics}\ }\textbf
  {\bibinfo {volume} {81}},\ \bibinfo {pages} {016401} (\bibinfo {year}
  {2017})}\BibitemShut {NoStop}%
\bibitem [{\citenamefont {Else}\ \emph {et~al.}(2020)\citenamefont {Else},
  \citenamefont {Monroe}, \citenamefont {Nayak},\ and\ \citenamefont
  {Yao}}]{DTCrev2}%
  \BibitemOpen
  \bibfield  {author} {\bibinfo {author} {\bibfnamefont {D.~V.}\ \bibnamefont
  {Else}}, \bibinfo {author} {\bibfnamefont {C.}~\bibnamefont {Monroe}},
  \bibinfo {author} {\bibfnamefont {C.}~\bibnamefont {Nayak}},\ and\ \bibinfo
  {author} {\bibfnamefont {N.~Y.}\ \bibnamefont {Yao}},\ }\bibfield  {title}
  {\bibinfo {title} {Discrete time crystals},\ }\href
  {https://doi.org/10.1146/annurev-conmatphys-031119-050658} {\bibfield
  {journal} {\bibinfo  {journal} {Annual Review of Condensed Matter Physics}\
  }\textbf {\bibinfo {volume} {11}},\ \bibinfo {pages} {467} (\bibinfo {year}
  {2020})}\BibitemShut {NoStop}%
\bibitem [{\citenamefont {Lamata}(2019)}]{sym11101310}%
  \BibitemOpen
  \bibfield  {author} {\bibinfo {author} {\bibfnamefont {L.}~\bibnamefont
  {Lamata}},\ }\bibfield  {title} {\bibinfo {title} {Symmetry in quantum optics
  models},\ }\bibfield  {journal} {\bibinfo  {journal} {Symmetry}\ }\textbf
  {\bibinfo {volume} {11}},\ \href
  {https://doi.org/https://doi.org/10.3390/books978-3-03921-859-2}
  {https://doi.org/10.3390/books978-3-03921-859-2} (\bibinfo {year}
  {2019})\BibitemShut {NoStop}%
\bibitem [{\citenamefont {Mangazeev}\ \emph {et~al.}(2021)\citenamefont
  {Mangazeev}, \citenamefont {Batchelor},\ and\ \citenamefont
  {Bazhanov}}]{sym2}%
  \BibitemOpen
  \bibfield  {author} {\bibinfo {author} {\bibfnamefont {V.~V.}\ \bibnamefont
  {Mangazeev}}, \bibinfo {author} {\bibfnamefont {M.~T.}\ \bibnamefont
  {Batchelor}},\ and\ \bibinfo {author} {\bibfnamefont {V.~V.}\ \bibnamefont
  {Bazhanov}},\ }\bibfield  {title} {\bibinfo {title} {The hidden symmetry of
  the asymmetric quantum rabi model},\ }\href
  {https://doi.org/10.1088/1751-8121/abe426} {\bibfield  {journal} {\bibinfo
  {journal} {Journal of Physics A: Mathematical and Theoretical}\ }\textbf
  {\bibinfo {volume} {54}},\ \bibinfo {pages} {12LT01} (\bibinfo {year}
  {2021})}\BibitemShut {NoStop}%
\bibitem [{\citenamefont {Li}\ and\ \citenamefont {Batchelor}(2021)}]{sym3}%
  \BibitemOpen
  \bibfield  {author} {\bibinfo {author} {\bibfnamefont {Z.-M.}\ \bibnamefont
  {Li}}\ and\ \bibinfo {author} {\bibfnamefont {M.~T.}\ \bibnamefont
  {Batchelor}},\ }\bibfield  {title} {\bibinfo {title} {Hidden symmetry and
  tunneling dynamics in asymmetric quantum rabi models},\ }\href
  {https://doi.org/10.1103/physreva.103.023719} {\bibfield  {journal} {\bibinfo
   {journal} {Physical Review A}\ }\textbf {\bibinfo {volume} {103}},\ \bibinfo
  {pages} {023719} (\bibinfo {year} {2021})}\BibitemShut {NoStop}%
\bibitem [{\citenamefont {Lu}\ \emph {et~al.}(2021)\citenamefont {Lu},
  \citenamefont {Li}, \citenamefont {Mangazeev},\ and\ \citenamefont
  {Batchelor}}]{sym4}%
  \BibitemOpen
  \bibfield  {author} {\bibinfo {author} {\bibfnamefont {X.}~\bibnamefont
  {Lu}}, \bibinfo {author} {\bibfnamefont {Z.-M.}\ \bibnamefont {Li}}, \bibinfo
  {author} {\bibfnamefont {V.~V.}\ \bibnamefont {Mangazeev}},\ and\ \bibinfo
  {author} {\bibfnamefont {M.~T.}\ \bibnamefont {Batchelor}},\ }\bibfield
  {title} {\bibinfo {title} {Hidden symmetry in the biased dicke model},\
  }\href {https://doi.org/10.1088/1751-8121/ac0f16} {\bibfield  {journal}
  {\bibinfo  {journal} {Journal of Physics A: Mathematical and Theoretical}\
  }\textbf {\bibinfo {volume} {54}},\ \bibinfo {pages} {325202} (\bibinfo
  {year} {2021})}\BibitemShut {NoStop}%
\bibitem [{\citenamefont {Drummond}\ and\ \citenamefont
  {Walls}(1980)}]{Drummond}%
  \BibitemOpen
  \bibfield  {author} {\bibinfo {author} {\bibfnamefont {P.~D.}\ \bibnamefont
  {Drummond}}\ and\ \bibinfo {author} {\bibfnamefont {D.~F.}\ \bibnamefont
  {Walls}},\ }\bibfield  {title} {\bibinfo {title} {Quantum theory of optical
  bistability. 1. nonlinear polarisability model},\ }\href
  {https://doi.org/10.1088/0305-4470/13/2/034} {\bibfield  {journal} {\bibinfo
  {journal} {Journal of Physics A: Mathematical and General}\ }\textbf
  {\bibinfo {volume} {13}},\ \bibinfo {pages} {725} (\bibinfo {year}
  {1980})}\BibitemShut {NoStop}%
\bibitem [{\citenamefont {Breuer}\ and\ \citenamefont
  {Petruccione}(2007)}]{Breuer2007}%
  \BibitemOpen
  \bibfield  {author} {\bibinfo {author} {\bibfnamefont {H.-P.}\ \bibnamefont
  {Breuer}}\ and\ \bibinfo {author} {\bibfnamefont {F.}~\bibnamefont
  {Petruccione}},\ }\href
  {https://doi.org/10.1093/acprof:oso/9780199213900.001.0001} {\emph {\bibinfo
  {title} {The Theory of Open Quantum Systems}}}\ (\bibinfo  {publisher}
  {Oxford University {PressOxford}},\ \bibinfo {year} {2007})\BibitemShut
  {NoStop}%
\bibitem [{\citenamefont {Risken}\ and\ \citenamefont
  {Vogel}(1988)}]{PhysRevA.38.1349}%
  \BibitemOpen
  \bibfield  {author} {\bibinfo {author} {\bibfnamefont {H.}~\bibnamefont
  {Risken}}\ and\ \bibinfo {author} {\bibfnamefont {K.}~\bibnamefont {Vogel}},\
  }\bibfield  {title} {\bibinfo {title} {Quantum tunneling rates in dispersive
  optical bistability for low cavity damping},\ }\href
  {https://doi.org/10.1103/PhysRevA.38.1349} {\bibfield  {journal} {\bibinfo
  {journal} {Phys. Rev. A}\ }\textbf {\bibinfo {volume} {38}},\ \bibinfo
  {pages} {1349} (\bibinfo {year} {1988})}\BibitemShut {NoStop}%
\bibitem [{\citenamefont {Anikin}\ \emph {et~al.}(2019)\citenamefont {Anikin},
  \citenamefont {Maslova}, \citenamefont {Gippius},\ and\ \citenamefont
  {Sokolov}}]{Anikin2019}%
  \BibitemOpen
  \bibfield  {author} {\bibinfo {author} {\bibfnamefont {E.~V.}\ \bibnamefont
  {Anikin}}, \bibinfo {author} {\bibfnamefont {N.~S.}\ \bibnamefont {Maslova}},
  \bibinfo {author} {\bibfnamefont {N.~A.}\ \bibnamefont {Gippius}},\ and\
  \bibinfo {author} {\bibfnamefont {I.~M.}\ \bibnamefont {Sokolov}},\
  }\bibfield  {title} {\bibinfo {title} {Enhanced excitation of a driven
  bistable system induced by spectrum degeneracy},\ }\href
  {https://doi.org/10.1103/PhysRevA.100.043842} {\bibfield  {journal} {\bibinfo
   {journal} {Phys. Rev. A}\ }\textbf {\bibinfo {volume} {100}},\ \bibinfo
  {pages} {043842} (\bibinfo {year} {2019})}\BibitemShut {NoStop}%
\bibitem [{\citenamefont {Dodonov}\ and\ \citenamefont
  {Man{'}ko}(2003)}]{mankododonov}%
  \BibitemOpen
  \bibfield  {author} {\bibinfo {author} {\bibfnamefont {V.~V.}\ \bibnamefont
  {Dodonov}}\ and\ \bibinfo {author} {\bibfnamefont {V.~I.}\ \bibnamefont
  {Man{'}ko}},\ }\href {https://doi.org/https://doi.org/10.1201/9781482288223}
  {\emph {\bibinfo {title} {Theory of Nonclassical States of Light}}}\
  (\bibinfo  {publisher} {CRC Press},\ \bibinfo {year} {2003})\BibitemShut
  {NoStop}%
\bibitem [{\citenamefont {Vogel}\ and\ \citenamefont
  {Risken}(1990)}]{vogelriskenbistable}%
  \BibitemOpen
  \bibfield  {author} {\bibinfo {author} {\bibfnamefont {K.}~\bibnamefont
  {Vogel}}\ and\ \bibinfo {author} {\bibfnamefont {H.}~\bibnamefont {Risken}},\
  }\bibfield  {title} {\bibinfo {title} {Dispersive optical bistability for
  large photon numbers and low cavity damping},\ }\href
  {https://doi.org/10.1103/physreva.42.627} {\bibfield  {journal} {\bibinfo
  {journal} {Physical Review A}\ }\textbf {\bibinfo {volume} {42}},\ \bibinfo
  {pages} {627} (\bibinfo {year} {1990})}\BibitemShut {NoStop}%
\bibitem [{\citenamefont {Maslova}\ \emph {et~al.}(2007)\citenamefont
  {Maslova}, \citenamefont {Johne},\ and\ \citenamefont
  {Gippius}}]{maslovagippiuskerr}%
  \BibitemOpen
  \bibfield  {author} {\bibinfo {author} {\bibfnamefont {N.~S.}\ \bibnamefont
  {Maslova}}, \bibinfo {author} {\bibfnamefont {R.}~\bibnamefont {Johne}},\
  and\ \bibinfo {author} {\bibfnamefont {N.~A.}\ \bibnamefont {Gippius}},\
  }\bibfield  {title} {\bibinfo {title} {Role of fluctuations in nonlinear
  dynamics of a driven polariton system in semiconductor microcavities},\
  }\href {https://doi.org/10.1134/s0021364007140123} {\bibfield  {journal}
  {\bibinfo  {journal} {{JETP} Letters}\ }\textbf {\bibinfo {volume} {86}},\
  \bibinfo {pages} {126} (\bibinfo {year} {2007})}\BibitemShut {NoStop}%
\end{thebibliography}%


%

\onecolumngrid
\appendix
\section{Non-degenerate perturbation theory corrections for the energy levels}
\label{Appendix_corrections}
In this section, we present the explicit expressions for the second- and the fourth-order non-degenerate perturbation theory corrections to the oscillator energy levels. The second-order perturbation correction for the level $n$ reads
\begin{equation}
	\label{eq:second_order_correction_expr}
g^2 \epsilon_n^{(2)} = \frac{|V_{n,n-1}|^2}{\epsilon_{n}^{(0)} - \epsilon_{n-1}^{(0)}} +
\frac{|V_{n,n+1}|^2}{\epsilon_{n}^{(0)} - \epsilon_{n+1}^{(0)}},
\end{equation}
where
\begin{equation} \label{pert}
V_{k,n} = g \left(\sqrt{n+1} \sqrt{N-n} \delta_{k, n+1}+\sqrt{n} \sqrt{N-n+1} \delta_{k, n-1}\right).
\end{equation}
For the two levels $n$ and $m-n$, the expressions \eqref{eq:second_order_correction_expr} are not explicitly symmetric. However, the contributions of the levels $n-1$ and $n+1$ have different signs and partially cancel each other. By algebraic manipulations, one can obtain 
\begin{equation} \label{corrections_en}
\begin{gathered}
\epsilon_{n}^{(2)}=\frac{1}{\alpha_1 + \alpha_2 } \frac{(2 n-\mu_N)^{2}-\mu_N^{2}+2 N(\mu_N+1)}{(2 n-\mu_N)^{2}-1},  \\
\end{gathered}
\end{equation}
which turns out to be symmetric with respect to the replacement $n \rightarrow \mu_N - n$. Much lengthier calculations also lead to the symmetric expression 
for the fourth-order correction
\begin{equation} \label{corrections_en_4}
\begin{gathered}
\epsilon_{n}^{(4)}=\frac{1}{(\alpha_1 + \alpha_2 )^{3}} \frac{N^{2} A+N B+C}{\left((2 n-\mu_N)^{2}-1\right)^{3}\left((2 n-\mu_N)^{2}-4\right)}, \\
A=-12(2 n-\mu_N)^{4}+4(2 n-\mu_N)^{2} \left(5 \mu_N^{2}+10 \mu_N+11\right)+  
4\left(7 \mu_N^{2}+14 \mu_N+4\right),  \\
B=12(2 n-\mu_N)^{4}(\mu_N-1)-4(2 n-\mu_N)^{2}\left(5 \mu_N^{3}+5 \mu_N^{2}+\mu_N-11\right)-
4\left(7 \mu_N^{3}+7 \mu_N^{2}-10 \mu_N-4\right),  \\
C=(2 n-\mu_N)^{6}-3(2 n-\mu_N)^{4} \left(2 \mu_N^{2}+3\right)+ 
(2 n-\mu_N)^{2} \left(5 \mu_N^{4}+2 \mu_N^{2}+20\right) +
7 \mu_N^{4}-20 \mu_N^{2}. 
\end{gathered}
\end{equation}

\section{The symmetry of the generalised Hamiltonian}
\label{D}
In this section, we prove the identity defined by Eq.~(12),
$\hat{\mathcal{H}}_{\nu, N}=\mathcal{T} I\hat{\mathcal{H}}_{\mu_N-\nu, N}I^{-1} \mathcal{T}^{-1}$ which proves the equivalence of 
two instances of the generalised oscillator Hamiltonians for the indices $\nu$ and $\mu_N - \nu$ (see Eqs.~(11) and (14)).

First of all, one can check that after the action of the isomorphism $I$, the right-hand-side of Eq.~(12) reads
\begin{multline} \label{h_nu_m}
  I\mathcal{\hat{H}}_{\mu_N - \nu, N}I^{-1} =\frac{1}{2}\left(\alpha_{1}+\alpha_{2}\right) \sum\limits_{\sigma - \nu \in \mathbb{Z}} \sigma (\sigma - \mu_N)\ket{\sigma} \bra{\sigma}  + \\
g \! \sum\limits_{\sigma - \nu \in \mathbb{Z}} \sqrt{(\mu_N-\sigma) (N - \mu_N +  \sigma + 1)} ( \ket{\sigma+1}\bra{\sigma} + \ket{\sigma}\bra{\sigma + 1} ).
\end{multline}

Second of all, let us define the new operators 
\begin{equation}
\hat{H}_1 = U^{-1}\mathcal{\hat{H}}_{\nu,N} U, \quad \hat{H}_2 = V^{-1}I\mathcal{\hat{H}}_{\mu_N-\nu, N} I^{-1}V.
\end{equation}
Using expressions (14) for $U$ and $V$, 
it easy to obtain that these operators can be written as 
\begin{equation}
\hat{H}_1 = \frac{1}{2}\left(\alpha_{1}+\alpha_{2}\right) \sum\limits_{\sigma - \nu \in \mathbb{Z}} \sigma (\sigma - \mu_N) \ket{\sigma} \bra{\sigma} + 
g\! \sum\limits_{\sigma - \nu \in \mathbb{Z}} \left( \sigma (N - \sigma + 1) \ket{\sigma-1}\bra{\sigma}  + \ket{\sigma}\bra{\sigma - 1}  \right), 
\end{equation}
\begin{equation}
\hat{H}_2 = \frac{1}{2}\left(\alpha_{1}+\alpha_{2}\right)\sum\limits_{\sigma - \nu \in \mathbb{Z}} \sigma (\sigma - \mu_N)\ket{\sigma} \bra{\sigma}  + 
g \!\sum\limits_{\sigma - \nu \in \mathbb{Z}} \left( (\mu_N-\sigma + 1) (N - \mu_N +  \sigma ) \ket{\sigma-1}\bra{\sigma}  + \ket{\sigma}\bra{\sigma - 1} \right).
\end{equation}
In terms of these operators, the identity of Eq.~(12) takes the form
\begin{equation} \label{213}
\hat{H}_1 T = T \hat{H}_2,
\end{equation}
Let us note that all the operators in Eq.~\eqref{213} depend on $g$. 

For $g = 0$, it is obvious that $\hat{H}_1 = \hat{H}_2$, and $T = \mathds{1}$. For non-zero $g$, let us search for the operator $T$ obeying Eq.~\eqref{213}
in the form of the Taylor series in $g$:
\begin{equation} 
  \label{214}
T = \mathds{1}  + \frac{2 g }{\left(\alpha_{1}+\alpha_{2}\right)} T^{(1)} + \left(\frac{2 g }{\left(\alpha_{1}+\alpha_{2}\right)} \right)^2T^{(2)} + \ldots .
\end{equation}
After substituting the expression \eqref{214} for $T$ 
into Eq.~\eqref{213}, one gets the identities for the matrix elements
$T_{\sigma,\sigma'}$ of the Taylor expansion coefficients of $T$:
\begin{multline}
  \label{eq:taylor_coef_identities}
\sigma(\sigma - \mu_N)T^{(k+1)}_{\sigma, \sigma'} + (\sigma + 1)(N - \sigma)T^{(k)}_{\sigma+1, \sigma'} +T^{(k)}_{\sigma - 1, \sigma'} = \\
\sigma'(\sigma' - \mu_N)T^{(k+1)}_{\sigma, \sigma'} + 
(\mu_N - \sigma' +1)(N - \mu_N + \sigma') T^{(k)}_{\sigma, \sigma'-1} + T^{(k)}_{\sigma, \sigma' + 1}.
\end{multline}
It may be easily proven by induction that the only 
non-zero matrix elements of the Taylor coefficients are 
$T^{(k)}_{\sigma,\sigma+k}$, and the values
\begin{equation}
T^{(k)}_{\sigma, \sigma+k} = \frac{(N-\mu_N) \ldots (N-\mu_N+k-1)}{k!} =  \frac{(N-\mu_N+k-1)!}{k!(N-\mu_N-1)!},
\end{equation}
indeed obey the identity \eqref{eq:taylor_coef_identities}. 
Also, let us note that
$T^{(k)}_{\sigma,\sigma+k}$ coincide with the Taylor coefficients of the function
\begin{equation}
\left(1 - x \right)^{-(N-\mu_N)} = \sum\limits_{k=0}^{\infty} {N-\mu_N+k-1 \choose k}{x^k}, \quad |x| < 1.
\end{equation}
So, the transformation operator $T$ can be cast to a more compact form
\begin{equation}
T = \left( \mathds{1} - \frac{2g}{\left(\alpha_{1}+\alpha_{2}\right) } \sum\limits_{\sigma - \nu \in \mathbb{Z}} \ket{\sigma} \bra{\sigma + 1} \right)^{- (N-\mu_N)}.
\end{equation}

\section{Energy splitting due to multi-photon resonance}
\label{E}
In this section, we calculate the matrix Green's function to demonstrate that the degeneracy in the spectrum of the model lifts at higher orders of perturbation theory. First of all, let us write the following matrix, containing the matrix element of the full Green's function
\begin{equation} \label{eq:matr_fun}
\mathcal{G}=\left[\begin{array}{ll}
G_{n,n} & G_{n, m-n} \\
G_{m-n, n} & G_{m-n, m-n}
\end{array}\right],
\end{equation}
here $G_{i,j} = \mel{i}{\hat{G}}{j}$ --- matrix element of the Green's function. The matrix function \eqref{eq:matr_fun} satisfy the following Dyson equation

\begin{equation} \label{Dysonn}
\mathcal{G} = \mathcal{G}^{(0)} + \mathcal{G}^{(0)} \Sigma \mathcal{G}.
\end{equation}

\begin{figure}[h!]
\center{\includegraphics[width=0.75\textwidth]{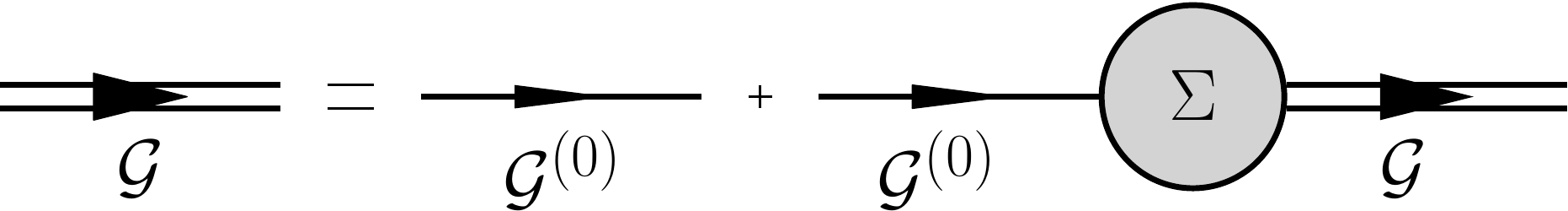}}
\caption{Feynman diagram for the Dyson equation for the matrix Green's function $\mathcal{G}$.}
\label{Diagram}
\end{figure}

This Dyson equation can be written in the form of the diagram, see Fig.~\ref{Diagram}.

In the equation \eqref{Dysonn} and in Fig.~\ref{Diagram}, we introduced the notation
\begin{equation} \label{ir}
\begin{gathered}
\mathcal{G}^{(0)}=\left[\begin{array}{cc}
\left(\omega-\epsilon_{n}^{(0)}\right)^{-1} & 0 \\
0 & \left(\omega-\epsilon_{m-n}^{(0)}\right)^{-1}
\end{array}\right], \quad
\Sigma = \begin{bmatrix}
\Sigma_{n,n}(\omega) & \Sigma_{n,m-n}(\omega) \\
\Sigma_{m-n,n}(\omega) & \Sigma_{m-n,m-n}(\omega)
\end{bmatrix}.
\end{gathered}
\end{equation}

In the formula \eqref{ir}, $\Sigma_{i,j}(\omega)$ it is a sum of all diagrams that start at $i$, finish at $j$ and do not contain the indices $i$ and $j$ in between.  In other words,  $\Sigma_{i,j}(\omega)$ has the following form
\begin{equation}
\label{irred}
\Sigma_{i,j}(\omega) = \sum\limits_{k_1}  \frac{V_{i,k_1}V_{k_1,j}}{\omega - \epsilon^{(0)}_{k_1}} + \sum\limits_{k_1,k_2} \frac{V_{i,k_1}V_{k_1, k_2}V_{k_2, j}}{(\omega - \epsilon^{(0)}_{k_1})(\omega - \epsilon^{(0)}_{k_2})} + \ldots + 
\sum\limits_{k_1, \ldots , k_{\ell}}\frac{V_{i,k_1} V_{k_1,k_2} \ldots V_{k_{\ell},  j}}{(\omega-\epsilon^{(0)}_{k_1}) \ldots (\omega - \epsilon^{(0)}_{k_{\ell}})} + \ldots .
\end{equation}

Due to the fact of Hermiticity of the considered Hamiltonian, irreducible terms are conjugate: $\Sigma_{n, m-n}(\omega)=\Sigma^{*}_{m-n, n}(\omega)$ (in our case they are real). One can write the formal solution of Dyson Eq.~(\ref{Dysonn})
\begin{equation} \label{Dyson_sol}
\begin{gathered}
\mathcal{G} = \left[\left( \mathcal{G}^{(0)} \right)^{-1} - \Sigma  \right]^{-1} = 
 \begin{bmatrix}
\omega - \epsilon_n^{(0)} - \Sigma_{n,n}(\omega) & -\Sigma_{n,m-n}(\omega) \\
-\Sigma_{m-n,n}(\omega) & \omega - \epsilon_{m-n}^{(0)} - \Sigma_{m-n,m-n}(\omega)
\end{bmatrix}^{-1}.
\end{gathered}
\end{equation}

Poles of (\ref{Dyson_sol}) are given by the following 
transcendental equation 
\begin{eqnarray}  \label{major}
\left(\omega - \epsilon^{(0)}_n - \Sigma_{n,n}(\omega)\right)\left(\omega -\epsilon^{(0)}_{m-n} - \Sigma_{m-n,m-n}(\omega)\right) =  |\Sigma_{n,m-n}(\omega)|^2.
\end{eqnarray}

To analyse the irreducible term (\ref{irred}), let us look at the matrix element of the perturbation part 
\begin{equation} \label{pert}
V_{k,n} = g \left(\sqrt{n+1} \sqrt{N-n} \delta_{k, n+1}+\sqrt{n} \sqrt{N-n+1} \delta_{k, n-1}\right).
\end{equation}
It is clear from (\ref{pert}) that non-vanishing terms 
satisfy $k = n \pm 1$. 

Using \eqref{major}, one can calculate the leading-order contribution
to the non-diagonal part of the self-energy matrix:
\begin{equation} \label{irredus}
\begin{gathered}
\Sigma_{n,m-n}(\omega) = 
\frac{V_{n,n+1}\dots V_{m-n-1, m-n}}{(\omega - \epsilon^{(0)}_{n+1}) \ldots (\omega - \epsilon^{(0)}_{m-n-1})} \propto g^{m-2n} .
\end{gathered}
\end{equation}

Now let us prove that the splitting between two eigenenergies 
at the integer values of $\mu_N$
indeed occurs only in the order $|m-2n|$ of $g$. For that, we will
use the symmetry of the perturbation theory corrections
discussed above (in Appendix~\ref{D}), and in Section~\ref{sec:symmetry_proof} of the main text.

The eigenenergies can be obtained from Eq.~\eqref{major}. This equation can be solved with help
of the perturbation theory expansions of $\Sigma_{i,j}(\omega)$.

One can see that the right part is equal to zero up to the 
$|m-2n|$-th order of $g$. So, let us first consider 
the equations
\begin{equation} \label{free_part}
\begin{gathered}
\omega - \epsilon^{(0)}_n - \Sigma_{n,n}(\omega) = 0, \\
\omega - \epsilon^{(0)}_{m-n} - \Sigma_{m-n,m-n}(\omega) = 0, 
\end{gathered}
\end{equation}
separately. Let us denote the roots of these equations as 
$\tilde\epsilon_1(\mu_N)$ and $\tilde\epsilon_2(\mu_N)$ respectively.
As $\Sigma_{i,j}(\omega)$ are regular in the vicinity of 
integer $\mu_N$, 
$\tilde\epsilon_1(\mu_N)$ and $\tilde\epsilon_2(\mu_N)$ are also 
regular and can be decomposed in power series in $g$.

We are interested in the degenerate or almost-degenerate case: in
this case, $\tilde\epsilon_1(\mu_N)$ and $\tilde\epsilon_2(\mu_N)$ 
are very close
to each other. It is convenient to consider the value of 
$\mu_N = \mu_N^*$ such as 
$\tilde\epsilon_1(\mu_{N}^{*}) = \tilde\epsilon_2(\mu_{N}^{*})$. For $g = 0$, 
$\mu_N^* = m$.

Let us now solve (\ref{major}) for the values of $\mu_N$ which
are close to $\mu_N^*$. For these values,
each expression in \eqref{free_part} can be 
replaced by $\omega - \tilde{\epsilon}_i$, $i = \{1, 2 \}$.
Also, the argument of $\Sigma_{n,m-n}(\omega)$ can be 
replaced by $(\tilde{\epsilon}_{1} + \tilde{\epsilon}_{2})/2$ 
in the leading order of $g$. 
After that, the transcendental equation \eqref{major} becomes 
quadratic, and its roots $\epsilon_{n,m-n}^\pm$ can be easily found
\begin{equation}
    \label{eq:energies_near_anticrossing}
    \epsilon_{n, m-n}^{\pm} \approx \frac{\tilde{\epsilon}_{1}+\tilde{\epsilon}_{2}}{2} \pm \sqrt{\left(\frac{\tilde{\epsilon}_{1}-\tilde{\epsilon}_{2}}{2}\right)^{2}+\left|\Sigma_{n, m-n}\right|^{2}},
\end{equation}
The energy spliting reads
\begin{equation}
    \label{eq:splitting_near_anticrossing}
    \epsilon_{n, m-n}^{+}-\epsilon_{n, m-n}^{-} \approx 2 \sqrt{\left(\frac{\tilde{\epsilon}_{1}-\tilde{\epsilon}_{2}}{2}\right)^{2}+\left|\Sigma_{n, m-n}\right|^{2}}. 
\end{equation}
The Eqs.~\eqref{eq:energies_near_anticrossing} and \eqref{eq:splitting_near_anticrossing} give the general 
leading-order expressions 
for the energies near the levels anti-crossing caused by 
multi-photon transition. 
According to Eq.~\eqref{eq:splitting_near_anticrossing}, 
the minimal level splitting 
is achieved at $\mu_N = \mu_N^*$.

Now we will apply these results for the considered model of two 
coupled nonlinear oscillators. Let us recall that this model 
has the symmetry of the non-degenerate perturbation theory 
corrections. This symmetry allows to show that the 
energy splitting at $\mu_N \in \mathds{Z}$ remains of order 
$\sim g^{m-2n}$ even though in general case 
$\mu_N^*$ depends on $g$.

Now let us write $\tilde\epsilon_1$ and $\tilde\epsilon_2$ in the form of power series in $g$
\begin{equation} \label{exp}
\begin{gathered}
\tilde\epsilon_1(\mu_N) = \epsilon^{(0)}_n + c_2 g^2 + c_4 g^4 + \ldots ,  \\
\tilde\epsilon_2(\mu_N) = \epsilon^{(0)}_{m-n} + d_2 g^2 + d_4 g^4 + \ldots .
\end{gathered}
\end{equation} 
Although $\tilde\epsilon_i(\mu_N)$ do not coincide exactly with the 
eigenenergies of the oscillator, their perturbation series coincide
with the non-degenerate perturbation series for $\epsilon_n(g)$ and 
$\epsilon_{m-n}(g)$ up to the order $2|m-2n|$. 
For $\tilde\epsilon_1(\mu_N)$, this can be verified by comparing 
the diagrammatic expansions for $\Sigma_{n,n}(\omega)$ and 
$\Sigma_n(\omega)$ (see Section~3). These 
expansion coincide up to the order $2|m-2n|$. Therefore, the same
is valid for $\tilde\epsilon_1$ and $\epsilon_n$, as the latter
is the solution of the equation 
$\omega - \epsilon_n^{(0)} - \Sigma_n(\omega) = 0$.

Therefore, in Eq.~(\ref{exp}) for the case 
$\mu_N \in \mathds{Z}$, 
all the coefficients $c_k$ and $d_k$ are equal 
to the non-degenerate perturbation theory corrections $ c_2 = d_2 = \epsilon_n^{(2)}$, $c_4 = d_4 = \epsilon_n^{(4)}$, etc up to the order $2|m-2n|$. As a result, at $\mu_N = m \in \mathds{Z}$, 
$\tilde\epsilon_1(\mu_N) - \tilde\epsilon_2(\mu_N) \sim g^{2(m-2n)}$.
Therefore, at the integer values of $\mu_N$, 
\begin{equation}
\epsilon_{n, m-n}^{+}-\epsilon_{n, m-n}^{-} = 
2\left|\Sigma_{n,m-n}\left(\epsilon_n^{(0)}\right)\right| + O\left(g^{m-2n+1}\right) \approx 
2\omega^R_{n,m-n},
\end{equation}
where $2\omega^R_{n,m-n}$ is the Rabi frequency of the multi-photon
transition between the states $\ket{n, N-n}$ and $\ket{m-n, N-m+n}$.
This means that the positions of anti-crossings indeed remain
unshifted as stated in Sections~1 and 3. More rigorously, one can prove that
$\mu_N^*(g) = m + O\left(g^{2(m-2n)+1}\right)$.

By substituting the actual values of the perturbation matrix elements
and the energies into the expression \eqref{irredus}, one can find

\begin{equation} \label{Rabi_fr}
\omega_{n, m-n}^{R}=\frac{1}{2}\left(\alpha_{1}+\alpha_{2}\right)\left(\frac{2 g}{\alpha_{1}+\alpha_{2}}\right)^{\! m-2 n} \sqrt{\frac{(m-n) !}{n !} \frac{(N-n) !}{(N-(m-n)) !}} \frac{1}{(m-2 n-1) !^{2}}.
\end{equation}

\section{The equilibrium points of the phase portrait and the critical coupling strength}
\label{B}
In this section, we find the critical value of the dimensionless coupling strength $\beta$ corresponding to the bifurcation on the phase portrait of 
two coupled oscillators. For that, we find the positions of the equilibrium points on the phase space from the equations of motion for $L_i$ and 
analyse their dependence on $g$.
From the Hamiltonian in terms of pseudo angular momentum operators (26), the equations of motion for the angular momenta can be found with help
of the Poisson brackets of $L_i$: $\{L_i,L_j\} = \epsilon_{ijk}L_k$ (which follow from the quantum--mechanical commutators of $\hat{L}_i$). The resulting 
equations of motion take the form
\begin{eqnarray}
\begin{dcases}
\frac{d {L}_{x}}{d t}=- (\alpha_1 + \alpha_2 ) {L}_{y} {L}_{z} - \frac{1}{2}\left( {\alpha_1 N} - {\alpha_2 N} - 2 \Delta \right) {L}_{y}, \\
\frac{d {L}_{y}}{d t}= (\alpha_1 + \alpha_2 ) {L}_{x} {L}_{z} + \frac{1}{2}\left( {\alpha_1 N} - {\alpha_2 N} - 2\Delta \right)  {L}_{x} - 2 g {L}_{z}, \\
\frac{d {L}_{z}}{d t}=2 g{L}_{y}.
\end{dcases}
\end{eqnarray}
Since we are looking for equilibrium points, all derivatives should be equal to zero. For all the equilibrium points, $L_y = 0$, therefore, they can be found
from the equation $dL_y/dt= 0$ together with the total pseudo momentum conservation law:
\begin{equation} \label{extr}
\begin{dcases}
  (\alpha_1 + \alpha_2 ) {L}_{x} {L}_{z} + \frac{1}{2}\left( {\alpha_1 N} - {\alpha_2 N} - 2\Delta \right)  {L}_{x} - 2 g {L}_{z} = 0,\\
  L_x^2 + L_z^2 = \frac{N(N+2)}{4} \approx (N/2)^2, \quad N \gg 1.
\end{dcases}
\end{equation}
The equilibrium points can be conveniently found with help of the
angular parametrisation for $L_x$ and $L_z$: \\$2L_z = {N}\cos{\vartheta}$, $2L_x = {N}\sin{\vartheta}$. The Eq.~\eqref{extr} reduce to the 
following trigonometric equation for $\vartheta$:
\begin{equation}
  \label{eq:stable_states_angles}
  \sin\vartheta + \left(1 - \frac{\mu_N}{N}\right) \tan{\vartheta} - \frac{4g}{N(\alpha_1 + \alpha_2)} = 0, \quad N \gg 1.
\end{equation}

At the critical value of $g$, two roots of this equation merge. However, a more convenient way to find it is not to explicitly examine the dependence of 
the roots of \eqref{eq:stable_states_angles} on $g$ but to consider formally the coupling constant $g$ as a function of $\vartheta$. 
The merging points of the roots 
correspond to the extrema of $g(\vartheta)$. 
By performing this procedure, one can obtain the following expression for the dimensionless coupling strength
$\beta = {8g^2N}/{(\alpha_1 + \alpha_2)^2\mu_N^3}$:
\begin{equation} \label{eq:beta_crit_app}
\beta_\mathrm{crit} = \frac{\gamma}{2}\left(\gamma^{\,2/3} - (\gamma-1)^{ 2/3}\right)^3,
\end{equation}
where $\gamma = N/\mu_N$.

\end{document}